\documentclass[sigconf,screen]{acmart}

\usepackage{verbatim}
\usepackage{color,colortbl}
\usepackage{xcolor}
\usepackage{soul}
\usepackage{graphicx}
\usepackage{booktabs} 
\usepackage{verbatim}
\usepackage{mdwlist}
\usepackage{listings}
\usepackage[linesnumbered,ruled,vlined]{algorithm2e}
\usepackage[english]{babel}
\usepackage[utf8]{inputenc}
\usepackage{amsmath}
\usepackage{enumitem}
\usepackage{amssymb}
\usepackage{url}
\usepackage{balance}

\definecolor{amethyst}{rgb}{0.6, 0.4, 0.8}

\definecolor{dkgreen}{rgb}{0,0.6,0}
\definecolor{gray}{rgb}{0.5,0.5,0.5}
\definecolor{mauve}{rgb}{0.58,0,0.82}

\setcopyright{acmcopyright}
\acmPrice{15.00}
\acmDOI{10.1145/3238147.3238215}
\acmYear{2018}
\copyrightyear{2018}
\acmISBN{978-1-4503-5937-5/18/09}
\acmConference[ASE '18]{Proceedings of the 2018 33rd ACM/IEEE International Conference on Automated Software Engineering}{September 3--7, 2018}{Montpellier, France}
\acmBooktitle{Proceedings of the 2018 33rd ACM/IEEE International Conference on Automated Software Engineering (ASE '18), September 3--7, 2018, Montpellier, France}

\begin{document}
%
\title[Empirically Assessing Opportunities for Prefetching and Caching in Mobile Apps]{Empirically Assessing Opportunities for\\ Prefetching and Caching in Mobile Apps}

\author{Yixue Zhao}
\affiliation{%
  \institution{University of Southern California}
  \city{Los Angeles}
  \state{California}
  \country{USA}
}
\author{Paul Wat}
\affiliation{%
  \institution{University of Southern California}
  \city{Los Angeles}
  \state{California}
  \country{USA}
}
\author{Marcelo Schmitt Laser}
\affiliation{%
  \institution{University of Southern California}
  \city{Los Angeles}
  \state{California}
  \country{USA}
}
\author{Nenad Medvidovi\'c}
\affiliation{%
  \institution{University of Southern California}
  \city{Los Angeles}
  \state{California}
  \country{USA}
}

\begin{abstract}
Network latency in mobile software has a large impact on user experience, with potentially severe economic consequences. Prefetching and caching have been shown effective in reducing the latencies in browser-based systems. However, those techniques cannot be directly applied to the emerging domain of mobile apps because of the differences in network interactions. Moreover, there is a lack of research on prefetching and caching techniques that may be suitable for the mobile app domain, and it is not clear whether such techniques can be effective or whether they are even feasible. This paper takes the first step toward answering these questions by conducting a comprehensive study to understand the characteristics of HTTP requests in over 1,000 popular Android apps. Our work focuses on the prefetchability of requests using static program analysis techniques and cacheability of resulting responses. We find that there is a substantial opportunity to leverage prefetching and caching in mobile apps, but that suitable techniques must take into account the nature of apps’ network interactions and idiosyncrasies such as untrustworthy HTTP header information. Our observations provide guidelines for developers to utilize prefetching and caching schemes in app development, and motivate future research in this area.
\end{abstract}



\maketitle

\section{Introduction}
\label{sec:intro}

There are over 5 billion mobile phone users and millions of mobile apps today~\cite{wearesocial}. The latency in mobile apps has been shown to have a large impact on  user experience and potentially severe economic consequences~\cite{wang2012far}. The main cause of user-perceived latency is the network, since the majority of mobile apps fetch data from the Internet regularly~\cite{ravindranath2012appinsight}.\footnote{In this context, we define latency as the response time of an HTTP request.}  Moreover, mobile devices rely
on wireless networks, which can exhibit intermittent
connectivity and low bandwidth~\cite{higgins2012informed}.

Optimizing network performance has long been studied in distributed systems, and prefetching and caching techniques have been shown as a high-reward way to reduce network latency: they can bypass the performance bottleneck (network speed) and mask latency by returning a response to a request from a local cache immediately~\cite{higgins2012informed}. 
While prefetching makes use of built-in caching schemes, caching-only techniques are also widely employed (e.g.,~\cite{zhang2013cachekeeper,qian2012web_ideal}). In this work, we study the two related phenomena separately: \textit{prefetchability} involves prefetching plus caching, while \textit{cacheability} involves only caching. 

The research on prefetching and caching techniques in the web browser domain has yielded a large body of work~\cite{wang2012far,mickens2010crom,netravali2016polaris,wang2013mobile,wang2016shandian,prefetchingclient,rosen2017push,prefetchingimpact}. However, the resulting techniques cannot be applied to mobile apps due to their different root causes of network latency. In the browser domain, the bottleneck for latency is resource loading since a large number of resources---usually files such as images---are needed within each  HTTP request~\cite{wang2011web}. In the mobile app domain, each request only fetches a single response, and additional requests need to be issued explicitly to fetch further resources~\cite{li2016automated,paloma_icse}. 
Thus, prefetching and caching techniques in the browser domain target subresources within a single request~\cite{rosen2017push,wang2012far,mickens2010crom,wang2016shandian}, while the research in the mobile app domain focuses on separate HTTP requests~\cite{zhang2013cachekeeper,paloma_icse}. 

Mobile users currently spend more than 80\% of their time in mobile apps, rather than using mobile browsers~\cite{appdominant}. Aside from a couple of exceptions, there has been a lack of research on prefetching and caching techniques that may be suitable for the mobile app domain. In fact, it is currently not clear whether such techniques can be effective or whether they are even feasible in practice. CacheKeeper~\cite{zhang2013cachekeeper} made an initial effort to study the redundant web traffic in mobile apps and proposed an OS-level caching service. However, the resulting  service was only evaluated on 10 apps. Furthermore, CacheKeeper's performance highly depends on the flaws in the web caching strategies employed in the original app, and its broader utility is unclear. Our previous work PALOMA~\cite{paloma_icse,mobilesoft2017src} used program analysis to identify HTTP requests that should be prefetched in mobile apps. We highlighted several program analysis challenges that can improve prefetching if addressed. However, PALOMA was evaluated on 32 apps. It is thus unclear to what extent PALOMA will be effective at a larger scale, and whether addressing the identified program analysis challenges is worthwhile. 

The dearth and shortcomings of previous work motivated us to conduct a more extensive empirical study  that aims to understand the characteristics of HTTP requests in mobile apps. In this paper, we report our results from the automated analysis of 1,687 most popular Android apps, spread across 33 app categories. Our work focuses on the prefetchability of requests (PALOMA's problem space) and cacheability of resulting responses (CacheKeeper's problem space). We found that a large number of HTTP requests used in real apps are prefetchable and the responses to those requests cacheable. This has the potential for significant reductions in user-perceived latency, which would, in turn, render the use of certain mobile apps even more attractive. 

At the same time, our study highlighted the need to carefully consider which requests should be prefetched and which data cached, for two reasons. First, we empirically demonstrated the frequent lack of discipline with which developers use the relevant HTTP headers in mobile (specifically, Android) apps, making those headers misleading. Second, we showed that responses to certain HTTP requests that seem like good candidates for caching may yield incorrect app behaviors due to cache staleness.

Our study is the first to provide extensive empirical evidence regarding the opportunities for prefetching and caching in mobile apps. It is also the first to identify concrete shortcomings in the current app development practices that are guaranteed to hinder solutions that may otherwise seem easy and intuitive. As a result, the study has the potential to motivate significant future research in this area. In this paper, we have identified several promising research directions.

The remainder of the paper is organized as follows. Section~\ref{sec:background} overviews the HTTP protocol and its use in the mobile app domain. Section~\ref{sec:rq} motivates and states our research questions. Section~\ref{sec:data} describes our collection and processing of the subject apps. Section~\ref{sec:result} discusses our findings and Section~\ref{sec:threat} describes the threats to their validity. A discussion of related work and conclusions round out the paper.

\section{background}
\label{sec:background}

In this section, we overview aspects of the  HTTP protocol that are relevant to prefetching and caching. 
We then illustrate with concrete examples of how developers perform network operations in mobile apps,  with a particular focus on Android. 
  
\subsection{HTTP Protocol}
\label{sec:sec:http_protocol}

Previous studies have shown that mobile apps spend between 34\% and 85\% of their time fetching data from the
Internet~\cite{ravindranath2012appinsight}. The majority of apps run over  HTTP~\cite{dai2013networkprofiler}, where requests are sent by clients and responses returned by servers. 

An \textit{HTTP request}  consists of an HTTP method, the destination of the resource to fetch (i.e., the URL), and request headers and body, both of which are optional. The HTTP method---\texttt{GET}, \texttt{POST}, \texttt{DELETE}, etc.---needs to be specified by developers when sending a request. 
Optional request headers allow the client to pass additional information to the server~\cite{rfc2616header}, such as \texttt{Accept-Language: en-US}. The request body contains the resource to send to the server, but is only needed for ``write'' HTTP methods, such as \texttt{POST}.

HTTP 1.1~\cite{rfc2616method} defines eight methods. Some of them, such as \texttt{DELETE}, are not suitable for prefetching because they may change the server's state contrary to the user's intention. Only the \texttt{GET} and \texttt{HEAD} methods are considered ``safe'', in that they result in the retrieval of data and do not have any side-effects on the server~\cite{rfc2616get,rfc2616safe}. The \texttt{HEAD} method is similar to \texttt{GET}, except that its response does not contain a message body~\cite{rfc2616get}. Thus, \texttt{GET} requests are of particular interest in our study. 

An \textit{HTTP response} consists of a status code, a status message, and response headers and body, both of which are optional. The status code and status message indicate whether the request was successful or not, and why. The response body contains the fetched resource from the server. Response headers contain additional information that is  often used by developers to decide on their caching strategies. For example, the \texttt{Expires} header specifies when the response will become stale, while \texttt{Cache-Control} header contains the information pertaining to caching mechanisms such as \texttt{no-cache} and \texttt{max-age}. Interestingly, as observed in our study (see Section~\ref{sec:result}), those headers  cannot always be trusted by developers, and sometimes they are missing altogether. 

\begin{lstlisting}[language=Java, basicstyle=\scriptsize\ttfamily, caption={Sending a POST request using the URLConnection library}, label={code:urlconnection}, captionpos=b]
  URL url = new URL("http://www.ase.com/post");
  URLConnection conn = url.openConnection();
  conn.setRequestMethod("POST"); 
  conn.setRequestProperty("Accept-Language", "en-US");
  OutputStreamWriter wr = new OutputStreamWriter(conn.getOutputStream()); 
  wr.write("post_data_to_send"); 
  wr.flush(); 
  InputStream responseStream = conn.getInputStream();
  Map headerMap = conn.getHeaderFields();
\end{lstlisting}

\begin{lstlisting}[language=Java, basicstyle=\scriptsize\ttfamily, caption={Sending a POST request using the OkHttp library}, label={code:okhttp}, captionpos=b]
  OkHttpClient client = new OkHttpClient();
  Request request = new Request.Builder()
                       .url("http://www.ase.com/post")
                       .addHeader("Accept-Language", "en-US")
                       .post("post_data_to_send")
                       .build();
  Response response = client.newCall(request).execute();
  Headers headers = response.headers();
\end{lstlisting}

\subsection{HTTP Libraries Used in Mobile Apps}
\label{sec:sec:http_library}
In Android apps, developers use off-the-shelf HTTP libraries to interact with servers. 
Listing~\ref{code:urlconnection} and Listing~\ref{code:okhttp} demonstrate how developers send HTTP requests and receive responses using the two most popular HTTP libraries for Android: \texttt{URLConnection} and \texttt{OkHttp}. 

When sending HTTP requests, developers need to specify the URL of the resource to be fetched (Listing~\ref{code:urlconnection}: line 1, Listing~\ref{code:okhttp}: line 3), HTTP method (Listing~\ref{code:urlconnection}: line 3, Listing~\ref{code:okhttp}: line 5), request headers (line 4 in both Listings), and request body (Listing~\ref{code:urlconnection}: line 6, Listing~\ref{code:okhttp}: line 5). Only the URL is mandatory and  \texttt{GET} method will be used by default if the HTTP method is not specified (e.g., if line 3 in Listing~\ref{code:urlconnection} and line 5 in Listing~\ref{code:okhttp} are removed). When receiving HTTP responses, developers can retrieve the response body (Listing~\ref{code:urlconnection}: line 8, Listing~\ref{code:okhttp}: line 7) as well as the response headers (Listing~\ref{code:urlconnection}: line 9, Listing~\ref{code:okhttp}: line 8) that may contain caching information.

\begin{figure*}[t!]
  \centering
  \includegraphics[width=\textwidth]{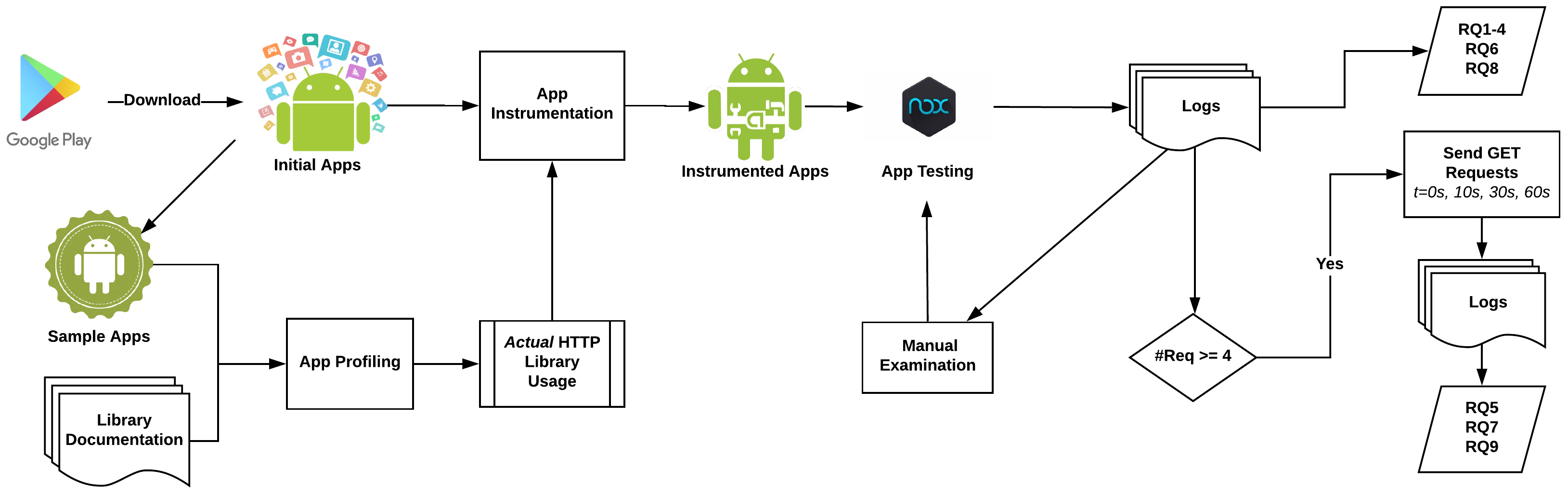}
  \vspace{-5mm}
  \caption{Our data collection workflow. The \emph{App Profiling}, \emph{App Instrumentation}, \emph{App Testing}, and \emph{Send GET Requests}   components perform automated tasks.}
  \label{fig:workflow}
\end{figure*}

\section{Research Questions}
\label{sec:rq}

The goal of this paper is to understand whether prefetching and caching can be applied to the mobile app domain  effectively, in order  to reduce user-perceived latency. We formulated nine research questions (RQs) to this end. These RQs target the \textit{prefetchability} of HTTP requests, \textit{cacheability} of HTTP responses, and \textit{redundancies} among HTTP requests.

\subsection{Prefetchability of HTTP Requests}
Our objective is to assess the extent to which requests in mobile apps are \textit{prefetchable}.  Prefetchable requests are read-only requests that have no side-effects on the server state. As discussed above (Section~\ref{sec:sec:http_protocol}), in the context of the HTTP protocol these are \texttt{GET} requests~\cite{rfc2616get}. Furthermore, we study whether the prevalence of prefetchable requests varies across different app categories. Such variations may allow identifying app categories that are particularly suitable for prefetching. 

We formulate three research questions to this end:

\begin{itemize}
\item \textbf{RQ$_1$} -- What is the number of \texttt{GET} requests per app?
\item \textbf{RQ$_2$} -- What is the percentage of \texttt{GET} requests among all HTTP requests in mobile apps? 
\item \textbf{RQ$_3$} -- How prevalent are \texttt{GET} requests across different app categories?
\end{itemize}

\subsection{Cacheability of HTTP Responses}
A prefetchable request may not be \textit{cacheable} if the response to the request changes over time (e.g., in the case of weather data). In such cases, the cached response may be stale and serving it would lead to incorrect app behavior. To determine when a response becomes stale, or whether a request is cacheable at all, developers have to rely on the header information specified in the response, specifically,  \texttt{Expires} and \texttt{Cache-Control}  (recall Section~\ref{sec:sec:http_protocol}). However, there are no standard rules for developers to follow when constructing a response, leaving open the possibility that header information may be unreliable or even missing. 

To investigate this, we formulate four additional research questions: 

\begin{itemize}
\item \textbf{RQ$_4$} -- How prevalent are \texttt{Expires} headers? 
\item \textbf{RQ$_5$} -- Are \texttt{Expires} headers trustworthy? \item \textbf{RQ$_6$} -- How prevalent are \texttt{Cache-Control} headers? 
\item \textbf{RQ$_7$} -- Are \texttt{Cache-Control} headers trustworthy?
\end{itemize}

\subsection{Identifying Truly Redundant HTTP Requests}
Caching is only effective when there exist redundant requests for the same resource. An HTTP request is \emph{redundant} if a previous request specified the same HTTP method and URL, 
and yielded the same response; the later request is redundant because the original response could have been stored locally and reused. Previous work~\cite{zhang2013cachekeeper} suggests an opportunity for mobile app-based caching techniques, in that it identified the presence of redundant HTTP traffic and showed that implementations of web caching are inadequate for mobile apps. 
Our work goes beyond identifying redundant HTTP requests and tries to assess the intent behind them. A set of \emph{ostensibly redundant} requests could be generated on purpose (e.g., to retrieve updated weather information), and thus may not be \emph{truly redundant}. If a caching scheme fails to consider this, it will lead to cache staleness. We thus consider the actual responses to the candidate redundant requests, aiming to distinguish among them and provide better insights for future caching techniques in mobile apps.

With this in mind, we formulate the last two research questions:
\begin{itemize}
\item \textbf{RQ$_8$} -- How prevalent are redundant HTTP requests?
\item \textbf{RQ$_9$} -- Are the identified ostensibly redundant requests truly redundant? 
\end{itemize}

\section{Data Collection}
\label{sec:data}

This section details (1) the workflow we used for data collection, (2) the criteria behind our selection of subject apps, (3) app instrumentation, (4) our collection of data via runtime testing, and (5) the reasons for eliminating certain apps from the subject set before conducting further analysis. All of the raw data regarding our subject apps and the corresponding code are publicly available~\cite{empirical_website}.

\subsection{Data Collection Workflow}
\label{sec:sec:workflow}
Figure~\ref{fig:workflow} illustrates the workflow we implemented for collecting the data needed to answer the nine research questions stated above. The initial subject apps were downloaded from the Google Play Store (Section~\ref{sec:sec:subject}). The apps were  automatically instrumented based on the information  extracted from HTTP library documentation and the decompiled code of several sample apps (Section~\ref{sec:sec:instrumentation}). The instrumented apps were automatically tested using randomly generated inputs to produce  logs that contain the information needed to answer RQ$_{1}$--RQ$_{4}$, RQ$_{6}$, and RQ$_{8}$ (Section~\ref{sec:sec:testing}). We manually examined the apps that could not be tested due to problems such as installation failures and runtime crashes, to identify the root causes of the problems (Section~\ref{sec:sec:final}). Finally, we automatically sent  \texttt{GET} requests to the subject apps at different time intervals, to answer RQ$_{5}$, RQ$_{7}$, and RQ$_{9}$ (Sections~\ref{sec:sec:cachability}~and~\ref{sec:sec:opportunity}).

\subsection{Initial Set of Subject Apps}
\label{sec:sec:subject}

We downloaded 1,687 top-ranked apps across 33 categories from the Google Play Store in the United States. 
1,308 of the apps could be processed by Soot~\cite{soot}, a state-of-the-art tool for instrumenting Android apps, as further discussed in Section~\ref{sec:sec:instrumentation}. 
The sizes of those 1,308 apps vary between 
16~KB and 103.4 MB. 
The total number of HTTP requests per app varied between 0 and 1,243 in our tests, as described in Section~\ref{sec:sec:testing}. 

Table~\ref{tbl:category} summarizes the information about the 1,308 subject apps. 
The table shows the maximum and average numbers of HTTP requests per app for each category; the minimum number of HTTP requests in every category is 0 and we thus omit it from the table. Finally, the right-most column shows the number of apps in each category that sent at least four HTTP requests in our tests, as well as the percentage of such apps compared to the total number of apps in the given category. The reason behind highlighting this subset of the 1,308 subject apps will be explained in Section~\ref{sec:sec:final}.

\subsection{App Instrumentation}
\label{sec:sec:instrumentation}

Each subject app went through an automated instrumentation process offline that used  Soot~\cite{soot} to insert code that captures  information about HTTP requests and responses. This information is primarily located in the HTTP headers. Capturing such information in the browser domain is straightforward because HTTP requests and responses are managed in a unified way. On the other hand, mobile apps presented a challenge: we first had to identify how the HTTP requests and responses are handled in different HTTP libraries (recall Section~\ref{sec:sec:http_library}); only then could we instrument the corresponding code to capture this information automatically. 

\begin{table}[b!]
\centering
\caption{App information for each category among initial subjects}
\vspace{-2mm}
\label{tbl:category}
\centering
 	\resizebox{\linewidth}{!}{
\begin{tabular}{|l|c|c|c|c|}
\hline
\textbf{Category}                 & \textbf{\#Apps} & \textbf{Max \#Req} & \textbf{Avg \#Req} & \textbf{\#Apps (\#Req$\geq$4)}  \\ \hline

~1. Art \& Design            & 11     & 14         & 2.27       & 3 (27.27\%)                      \\ \hline
~2. Auto \& Vehicles         & 29     & 6          & 1.07       & 4 (13.79\%)                      \\ \hline
~3. Beauty                   & 11     & 1243       & 120.82     & 6 (54.55\%)                      \\ \hline
~4. Books \& Reference       & 40     & 108        & 11.58      & 16 (40\%)                        \\ \hline
~5. Business                 & 55     & 87         & 5.71       & 17 (30.91\%)                     \\ \hline
~6. Comics                   & 55     & 319        & 20.84      & 19 (34.55\%)                     \\ \hline
~7. Communications           & 40     & 96         & 3.98       & 8 (20\%)                         \\ \hline
~8. Dating                   & 16     & 334        & 29.94      & 6 (37.5\%)                       \\ \hline
~9. Education                & 55     & 62         & 4.98       & 17 (30.91\%)                     \\ \hline
10. Entertainment            & 28     & 134        & 12.36      & 11 (39.29\%)                     \\ \hline
11. Events                   & 8      & 53         & 14.13      & 5 (62.5\%)                       \\ \hline
12. Finance                  & 61     & 150        & 15.97      & 27 (44.26\%)                     \\ \hline
13. Food \& Drink            & 28     & 188        & 16.43      & 13 (46.43\%)                     \\ \hline
14. Games                    & 37     & 59         & 12.59      & 25 (67.57\%)                     \\ \hline
15. Health \& Fitness        & 41     & 14         & 3.44       & 15 (36.59\%)                     \\ \hline
16. House \& Home            & 25     & 149        & 17.96      & 8 (32\%)                         \\ \hline
17. Libraries \& Demo        & 45     & 22         & 0.6        & 1 (2.22\%)                       \\ \hline
18. Lifestyle                & 21     & 82         & 12.48      & 12 (57.14\%)                     \\ \hline
19. Maps \& Navigation       & 54     & 206        & 8.37       & 8 (14.81\%)                      \\ \hline
20. Medical                  & 59     & 63         & 2.8        & 10 (16.95\%)                     \\ \hline
21. Music \& Audio           & 43     & 44         & 5.47       & 14 (32.56\%)                     \\ \hline
22. News \& Magazines        & 49     & 802        & 37.71      & 26 (53.06\%)                     \\ \hline
23. Parenting                & 24     & 28         & 2.54       & 5 (20.83\%)                      \\ \hline
24. Personalization          & 31     & 288        & 29.61      & 11 (35.48\%)                     \\ \hline
25. Photography              & 43     & 58         & 7.72       & 14 (32.56\%)                     \\ \hline
26. Productivity             & 68     & 119        & 8.31       & 24 (35.29\%)                     \\ \hline
27. Shopping                 & 46     & 198        & 21.54      & 22 (47.83\%)                     \\ \hline
28. Social                   & 48     & 108        & 10.4       & 23 (47.92\%)                     \\ \hline
29. Sports                   & 43     & 146        & 19.42      & 18 (41.86\%)                     \\ \hline
30. Tools                    & 54     & 130        & 6.44       & 16 (29.63\%)                     \\ \hline
31. Travel \& Local          & 63     & 208        & 14.33      & 27 (42.86\%)                     \\ \hline
32. Video Players \& Editors & 47     & 134        & 5.89       & 8 (17.02\%)                      \\ \hline
33. Weather                  & 30     & 123        & 14.7       & 12 (40\%)                        \\ \hline
\textbf{Total}                    & \textbf{1308}   & \textbf{1243}       & \textbf{50.20}      & \textbf{451 (34.48\%)}                    \\ \hline
\end{tabular}
}
\end{table}

It was thus necessary to determine what libraries most apps  use to send HTTP requests. We first identified a set of popular HTTP libraries, including \texttt{URLConnection}~\cite{urlconnection}, \texttt{OkHttp}~\cite{okhttp}, \texttt{Volley}~\cite{volley}, and \texttt{Retrofit}~\cite{retrofit}. We then analyzed a sample of the subject apps' bytecodes and checked the package names against the libraries. For example, the presence of the string ``java.net.URLConnection'' generally indicates the use of the \texttt{URLConnection} library. 

The data gathered from our analysis point to \texttt{URLConnection} and \texttt{OkHttp} as the most popular HTTP libraries used in the subject apps. This is unsurprising: \texttt{URLConnection}  is the standard built-in library of the Android framework, and it has been augmented with \texttt{OkHttp}  since Android v.4.4 (KitKat). We thus decided to focus on \texttt{URLConnection} and \texttt{OkHttp} in our study. 

We then performed a more detailed analysis of how our subject apps  use these two libraries. We recorded the runtimes of those methods that are imported from \texttt{URLConnection} and \texttt{OkHttp}, and narrowed our focus to methods that are most time-consuming. The rationale is that those are most likely to be the methods related to sending requests and receiving responses over the network. 

In addition, we inspected the decompiled code of the subject apps, as well as the documentation and source code of the HTTP libraries used in the apps, to identify the actual usage of HTTP requests and responses. The reason for this additional inspection is that developers send requests and receive responses in various ways, even when using the same HTTP library. Listings~\ref{code:urlconnection}~and~\ref{code:okhttp} in Section~\ref{sec:background} only demonstrate one common way of using each of the two HTTP libraries. While recommended in the libraries' documentation, there is no requirement or guarantee that developers will follow this guidance in their apps. Furthermore, the examples in the documentation are at the source code level, while our instrumentation using Soot~\cite{soot} is at the bytecode level. This meant that we needed to understand the actual usage of those two HTTP libraries at the bytecode level. With the additional inspection, we were able to identify the actual methods used for sending requests and receiving responses in the apps,  allowing us to instrument the code to capture the precise information needed for our study. For example, line 9 in Listing~\ref{code:urlconnection} 
defines \texttt{headerMap} that contains all of the header information; our instrumentation then inserts a method after line 9 to capture the headers relevant to our study, such as \texttt{Expires} header. It is important to note that the instrumented apps' primary functionality is left unchanged in this process. 

\subsection{App Testing}
\label{sec:sec:testing}

After the instrumentation, each app was subjected to random input testing through Android Debug Bridge (adb)~\cite{adb}. We used the UI/Application exerciser tool Monkey~\cite{monkey} to generate random streams of user events, such as clicks, touches, and swipes. We used random events in this study for two reasons: (1) to avoid bias introduced by particular user behaviors and (2) to generate large volumes of runtime requests automatically, which would not be practical if we relied on a human user. This is further discussed in Section~\ref{sec:threat}. The apps were run on the NoxPlayer Android emulator~\cite{noxplayer}. Each test consisted of 3,000 events under WiFi network settings. We also explored testing with 1,000, 5,000, and 10,000 events. We found that 3,000 was the smallest number of events that yielded a representative number of HTTP requests triggered at runtime across the subject apps; neither 5,000 nor 10,000 events resulted in a significant increase in HTTP requests, while 1,000 events proved to be too few to adequately exercise the relevant functionality in the apps. 

All tests were preceded by a fresh installation of the given subject app, and the app was removed from the emulator after each test's conclusion. This minimized the chances of errors caused by any interference between apps or by previously saved settings.

\subsection{Final Set of Subject Apps}
\label{sec:sec:final}

The objective of our study is to determine whether and when HTTP requests should be prefetched and their responses cached. In some cases, the number of  HTTP requests triggered in our tests was very low, suggesting that prefetching and caching in such apps would not be beneficial. To determine the nature of ``low network usage'' apps and the underlying reasons behind the data we obtained, we manually inspected each app, starting with those that do not trigger any requests. 

A total of 623 out of the 1,308 subject apps triggered no requests. We identified six recurring reasons behind this:
\begin{enumerate}
\item The app's installation failed.
\item The app crashed upon launching.
\item The app's version was incompatible with the NoxPlayer Android emulator~\cite{noxplayer}.
\item The app was obfuscated so that the methods relevant to HTTP requests were not captured by our instrumentation.
\item The app required external information before it could be used, such as a bank PIN (commonly required in the \emph{Finance} category) or a vehicle license plate (commonly required in the \emph{Auto \& Vehicles} category). 
\item The app only contained static content and did not rely on the network.
\end{enumerate}
Note that, while we could not automatically test the above apps, many of them may, in fact, trigger HTTP requests at runtime. The only exception are apps from the last category. The automated nature of our app testing prevented us from determining the exact numbers of apps that fell in each of the above six categories. A manual inspection of a random sample of the apps suggests that, with a 95\% confidence level, no more than 50\% of the 623 apps contained only static content. 


An additional 234 of the 1,308 subject apps triggered 1-3 requests at runtime. We observed a common pattern among these apps. Namely, regardless of the type of app, those requests tended to be one or more of the following: 
\begin{enumerate}
\item Load an application-specific configuration file.
\item Log in with Facebook using Facebook \texttt{GraphRequest}.
\item Use monitoring services, such as \texttt{Crashlytics} or \texttt{Google Analytics}. 
\end{enumerate}
Further manual testing of these apps yielded no additional HTTP requests beyond the above three. This finding shows a common usage of popular third-party services in mobile app development, whose impact on app performance should also be taken into account in terms of overhead, data usage, and energy consumption.


We were unable to identify any patterns such as the above  in apps that trigger any other number of requests. Thus, the below analysis of \textit{prefetchability} and \textit{cacheability} is based on 451 of our subject apps that trigger four or more requests at runtime, corresponding to the right-most column of Table~\ref{tbl:category}.

\section{Results and Discussion}
\label{sec:result}

This section describes the results of our analysis, framed by the nine research questions from Section~\ref{sec:rq}, and discusses the lessons learned from the results. Table~\ref{tbl:final_apps} summarizes the information about the final set of 451 subject apps in each category that are analyzed in this section. Note that the app categories are numbered 1-33, to aid the depiction and understanding of the figures in the remainder of this section. Among the 451 apps, the number of HTTP requests ranged between 4 (the cut-off number for our analysis, as discussed above) and 1,243, with the average slightly above 35 requests per app. 


\subsection{Prefetchability of HTTP Requests}
\label{sec:sec:prefetchability}

Recall from Section~\ref{sec:rq} that we try to answer three research questions regarding the prefetchability of HTTP requests. Specifically, we are interested in \texttt{GET} requests, which are the primary candidates for prefetching.
\begin{itemize}
\item \textbf{RQ$_1$} -- What is the number of \texttt{GET} requests per app?
\item \textbf{RQ$_2$} -- What is the percentage of \texttt{GET} requests among all HTTP requests in mobile apps? 
\item \textbf{RQ$_3$} -- How prevalent are \texttt{GET} requests across different app categories?
\end{itemize}

\begin{figure}[b!]
	\centering
		\includegraphics[width=0.48\textwidth]{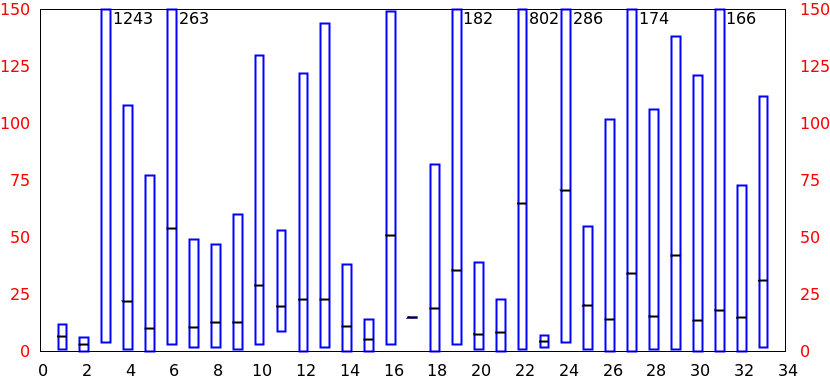}
        \vspace{-4mm}
	\caption{Minimum (bottom edges), maximum (top edges), and average (horizontal dashes) \emph{numbers} of \texttt{GET} requests in apps across the 33 app categories. Apps in 7 categories had maximums higher than 150 (numbers displayed beside the corresponding bars). Note that the average for app category 3 is also higher than 150, and thus not shown.} 
	\label{fig:get_number}
\end{figure}

\begin{figure}[b!]
	\centering
		\includegraphics[width=0.48\textwidth]{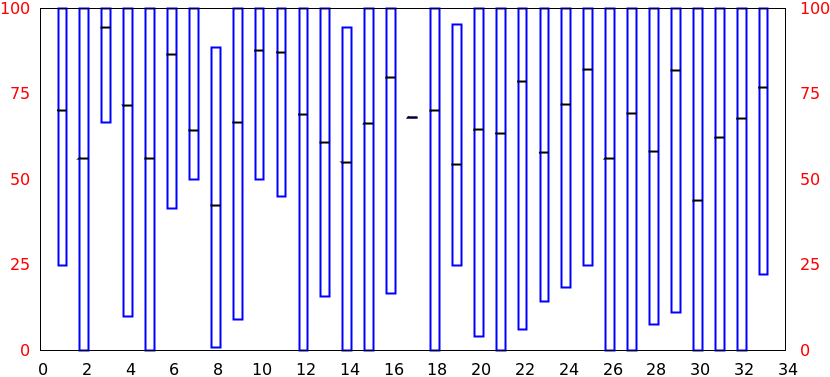}
        \vspace{-4mm}
	\caption{Minimum (bottom edges), maximum (top edges), and average (horizontal dashes) \emph{percentages} of \texttt{GET} requests in apps across the 33 app categories.}
	\label{fig:get_percentage}
\end{figure}

To answer the above questions, we instrumented and tested our subject apps using the procedure described in Section~\ref{sec:data}. We calculated the total number of \texttt{GET} requests observed during our testing, and the percentage of \texttt{GET} requests among all HTTP requests triggered at runtime in each app. We subsequently grouped the results by app category. 
Figure~\ref{fig:get_number} depicts the minimum, maximum, and average \emph{numbers} of \texttt{GET} requests per app (\emph{RQ$_1$})  across the different categories~(\emph{RQ$_3$}). Figure~\ref{fig:get_percentage} depicts the  minimum, maximum, and average \emph{percentages} of \texttt{GET} requests as compared to all HTTP requests~(\emph{RQ$_2$}) in each app category (\emph{RQ$_3$}). 

Our data indicate that \texttt{GET} requests are pervasive across all 33 app categories. As shown in Figure~\ref{fig:get_number}, seven categories contained apps that sent 150 or more \texttt{GET} requests. On average, an app sent 28 \texttt{GET} requests, and those requests comprised 68\% of all HTTP requests sent by the app.  As shown in Figure~\ref{fig:get_percentage}, several categories---\emph{Beauty} (94\%), \emph{Comics} (87\%), \emph{Entertainment} (88\%), and \emph{Events} (87\%)---had very high percentages of \texttt{GET} requests. Only two categories---\emph{Dating} (43\%) and \emph{Tools} (44\%)---had slightly fewer than 50\% of \texttt{GET} requests.

These results suggest that there is a significant opportunity to exploit prefetching among the 451 subject apps that sent 4 or more HTTP requests. It was surprising to see that 102 apps, spanning 29 of the 33 categories, sent only \texttt{GET} requests. Certain categories are potentially more suitable for prefetching than others. This is a by-product of the types of functionality that are typical in a given category. The nature of apps in ``stable'' domains, such as \emph{Art \& Design} or \emph{Libraries \& Demo}, is such that they may be able to operate with less remotely accessed data than apps in more ``dynamic'' domains such as \emph{News \& Magazines} or \emph{Shopping}. This  suggests that prefetching and caching techniques may benefit from leveraging knowledge regarding an app's domain.

\begin{table}[hbt!]
\centering
\caption{App information for each category among final subjects}
\label{tbl:final_apps}
\centering
\resizebox{\linewidth}{!}{

\begin{tabular}{|l|c|c|c|c|}
\hline
\textbf{Category}   & \textbf{\#Apps} & \textbf{Min. \#Req} & \textbf{Max. \#Req} & \textbf{Avg. \#Req} \\ \hline
~1. Art \& Design      & 3               & 4                   & 14                  & 8.33                \\ \hline
~2. Auto \& Vehicles   & 4               & 4                   & 6                   & 4.75                \\ \hline
~3. Beauty              & 6               & 4                   & 1243                & 220.33              \\ \hline
~4. Books \& Reference & 16              & 4                   & 108                 & 27.94               \\ \hline
~5. Business            & 17              & 4                   & 87                  & 17.24               \\ \hline
~6. Comics              & 19              & 4                   & 319                 & 59.58               \\ \hline
~7. Communications       & 8               & 4                   & 96                  & 19                  \\ \hline
~8. Dating              & 6               & 5                   & 334                 & 78.83               \\ \hline
~9. Education           & 17              & 4                   & 62                  & 15.06               \\ \hline
10. Entertainment       & 11              & 6                   & 134                 & 30.73               \\ \hline
11. Events              & 5               & 11                  & 53                  & 22.2                \\ \hline
12. Finance             & 27              & 5                   & 150                 & 35.59               \\ \hline
13. Food \& Drink      & 13              & 4                   & 188                 & 33.46               \\ \hline
14. Games              & 25              & 4                   & 59                  & 18                  \\ \hline
15. Health \& Fitness  & 15              & 4                   & 14                  & 8.13                \\ \hline
16. House \& Home      & 8               & 4                   & 149                 & 55.38               \\ \hline
17. Libraries \& Demo  & 1               & 22                  & 22                  & 22                  \\ \hline
18. Lifestyle           & 12              & 4                   & 82                  & 21                  \\ \hline
19. Maps \& Navigation & 8               & 8                   & 206                 & 54.88               \\ \hline
20. Medical             & 10              & 4                   & 63                  & 14                  \\ \hline
21. Music \& Audio      & 14              & 5                   & 44                  & 16.14               \\ \hline
22. News \& Magazines  & 26              & 4                   & 802                 & 70.88               \\ \hline
23. Parenting           & 5               & 4                   & 28                  & 12                  \\ \hline
24. Personalization     & 11              & 6                   & 288                 & 82.73               \\ \hline
25. Photography         & 14              & 4                   & 58                  & 23                  \\ \hline
26. Productivity        & 24              & 4                   & 119                 & 22.67               \\ \hline
27. Shopping            & 22              & 4                   & 198                 & 44.14               \\ \hline
28. Social              & 23              & 4                   & 108                 & 20.35               \\ \hline
29. Sports              & 18              & 7                   & 146                 & 45.67               \\ \hline
30. Tools               & 16              & 4                   & 130                 & 21.44               \\ \hline
31. Travel \& Local    & 27              & 4                   & 208                 & 32.11               \\ \hline
32. Video Players \& Editors       & 8               & 4                   & 134                 & 33.63               \\ \hline
33. Weather             & 12              & 7                   & 123                 & 36.17               \\ \hline
\textbf{Total}               & \textbf{451}          & \textbf{4}                & \textbf{1243}             & \textbf{35.28}               \\ \hline
\end{tabular}

}
\end{table}

\subsection{Cacheability of HTTP Responses}
\label{sec:sec:cachability}

As discussed in Section~\ref{sec:rq}, the cacheability of HTTP responses is a function of the presence of \texttt{Cache-Control} and \texttt{Expires} headers, and their trustworthiness. To that end, we try to answer the following four research questions. 
\begin{itemize}
\item \textbf{RQ$_4$} -- How prevalent are \texttt{Expires} headers? 
\item \textbf{RQ$_5$} -- Are \texttt{Expires} headers trustworthy? 
\item \textbf{RQ$_6$} -- How prevalent are \texttt{Cache-Control} headers? 
\item \textbf{RQ$_7$} -- Are \texttt{Cache-Control} headers trustworthy?
\end{itemize}

To answer the above questions, we instrumented the subject apps to capture response headers (recall Section~\ref{sec:sec:instrumentation}) and calculate the numbers of occurrences of the two relevant headers. 
To determine whether the header of a given request is trustworthy, we made each request 4 times: at initial time $t$, $t+10$, $t+30$, and $t+60$ seconds. This allowed us to determine whether later responses reflect what is specified in the header of the original response.

\begin{figure}[t!]
	\centering
		\includegraphics[width=0.48\textwidth]{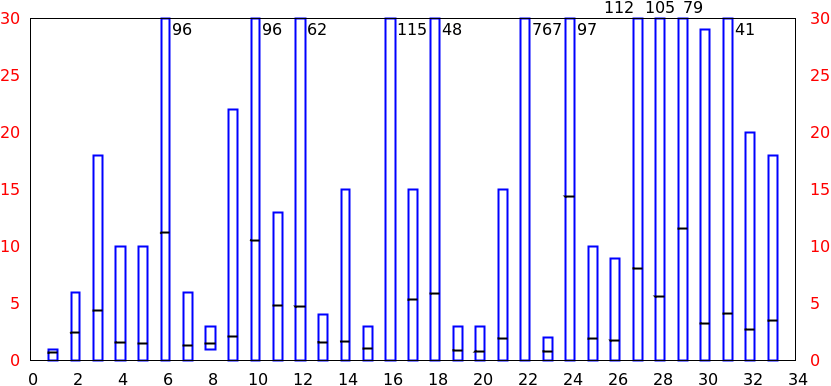}
        \vspace{-4mm}
	\caption{Minimum (bottom edges), maximum (top edges), and average (horizontal dashes) \emph{numbers} of \texttt{Expires} headers in each app category. Apps in 11 categories had maximums higher than 30 (numbers displayed beside or above the corresponding bars).}
	\label{fig:expires_number}
\end{figure}

\begin{figure}[t!]
	\centering
		\includegraphics[width=0.48\textwidth]{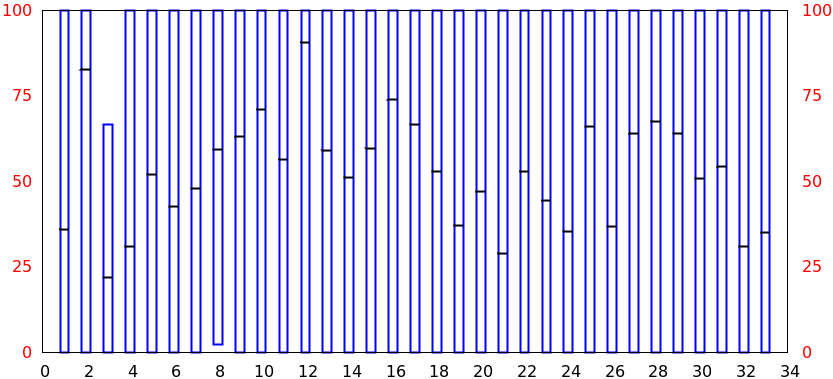}
        \vspace{-4mm}
	\caption{Minimum (bottom edges), maximum (top edges), and average (horizontal dashes) \emph{percentages} of the \texttt{Expires} headers for each app category.}
	\label{fig:expires_percentage}
\end{figure}

\begin{figure}[t!]
	\centering
		\includegraphics[width=0.48\textwidth]{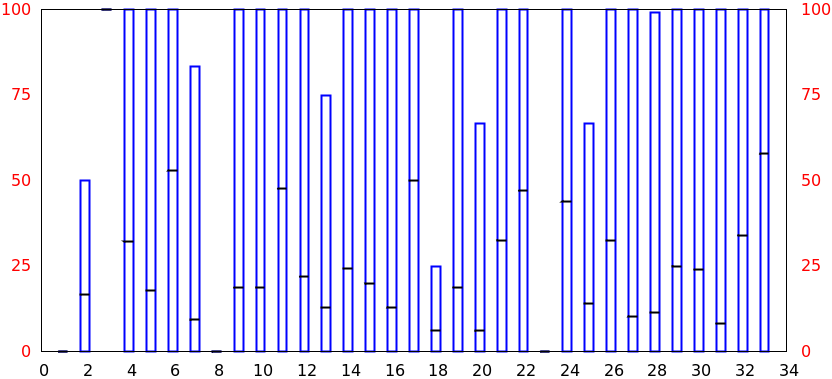}
	        \vspace{-4mm}
\caption{Minimum (bottom edges), maximum (top edges), and average (horizontal dashes) percentages of \emph{trusted}  \texttt{Expires} headers in each app category.}
	\label{fig:expires_trust}
\end{figure}

\begin{figure}[t]
	\centering
		\includegraphics[width=0.48\textwidth]{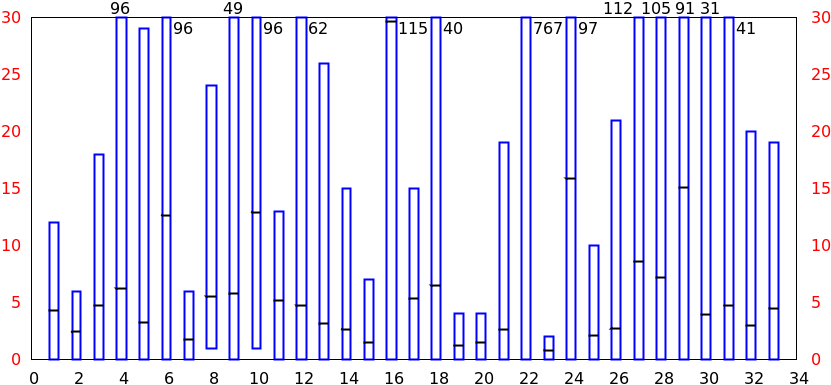}
        \vspace{-4mm}
	\caption{Minimum (bottom edges), maximum (top edges), and average (horizontal dashes) \emph{numbers} of \texttt{Cache-Control} headers in each app category. Apps in 14 categories had maximums higher than 30 (numbers displayed beside or above the corresponding bars).}
	\label{fig:cache_number}
\end{figure}

\begin{figure}[t!]
	\centering
		\includegraphics[width=0.48\textwidth]{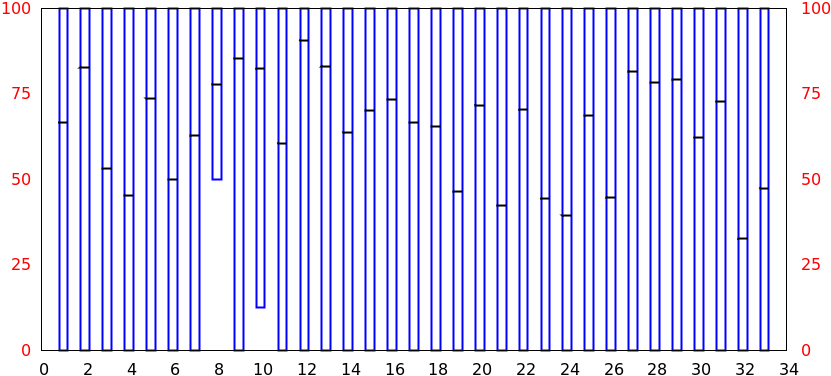}
        \vspace{-4mm}
	\caption{Minimum (bottom edges), maximum (top edges), and average (horizontal dashes)   \emph{percentages} of \texttt{Cache-Control} headers in each app category.}
	\label{fig:cache_percentage}
\end{figure}

\begin{figure}[t!]
	\centering
		\includegraphics[width=0.48\textwidth]{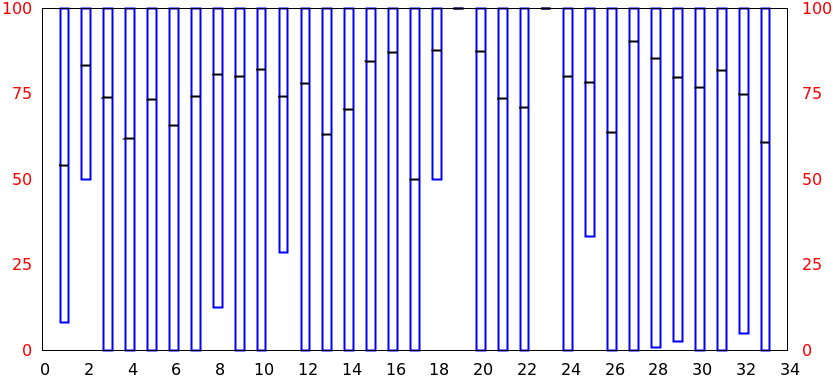}
        \vspace{-4mm}
	\caption{Minimum (bottom edges), maximum (top edges), and average (horizontal dashes) percentages of \emph{trusted} \texttt{Cache-Control} headers in each app category.}
	\label{fig:cache_trust}
\end{figure}

For example, let us assume that the original request is sent at \texttt{time(t)} and that the response header contains \texttt{Expires:~time(exp)}. We will mark the header as untrustworthy if it falls into any of the following three cases, where \texttt{x} is the time period after the original request is sent: 
\begin{enumerate}
\item \texttt{time(exp)} \textless= \texttt{time(t)}
\item (\texttt{time(t)} \textless  ~\texttt{time(exp)} $\leq$ \texttt{time(t+x)})\\ $\land$ (\texttt{response@(t)} = \texttt{response@(t+x)}) 
\item (\texttt{time(exp)} \textgreater ~\texttt{time(t+x)})\\ $\land$  
(\texttt{response@(t)} $\neq$ \texttt{response@(t+x)})
\end{enumerate}
In our case, \texttt{x} is any of $10s$, $30s$, or $60s$. The first case indicates a scenario where the response expires before the request is even sent. The second case indicates a scenario where the response is supposed to have expired, but it has remained unchanged. Finally, the third case indicates a scenario where the response should have remained the same, but it changed.

We use the analogous algorithm to determine whether the \texttt{Cache-} \texttt{Control} header is trustworthy, based on the \texttt{max-age} field specified within the header.

Figure~\ref{fig:expires_number} shows the minimum, maximum, and average numbers of the  \texttt{Expires} headers included in HTTP responses for each app category~(\emph{RQ$_4$}). 
Figure~\ref{fig:expires_percentage} shows the minimum, maximum, and average percentages of the \texttt{Expires} headers among all the response headers in each app category~(\emph{RQ$_4$}). Figure~\ref{fig:expires_trust} shows the percentages of the trustworthy \texttt{Expires} headers among all the \texttt{Expires} headers~(\emph{RQ$_5$}).   Figures~\ref{fig:cache_number},~\ref{fig:cache_percentage}, and~\ref{fig:cache_trust} show the analogous information for the \texttt{Cache-Control} header~(\emph{RQ$_6$}, \emph{RQ$_7$}).

From the results, we can conclude that the \texttt{Expires} headers and \texttt{Cache-Control} headers are not always included in the responses, and they are not always trustworthy. The \texttt{Cache-Control} header tends to be used more reliably than the \texttt{Expires} header. 
Across the 33 app categories, 53\% of the response headers contain \texttt{Expires} on average, while 65\% contain \texttt{Cache-Control}. Only an average of 25\% of the \texttt{Expires} headers are trustworthy, while 77\% of the \texttt{Cache-Control} headers are trustworthy. While there are individual apps among our subjects where each of the two headers was used in a completely trustworthy manner (100\%), there were an even greater number of apps where the opposite was true (0\%).

These results strongly suggest that developers should not depend on the response headers to determine their caching schemes. Unfortunately, there are currently no reliable alternatives for the mobile app domain. However, this presents a research opportunity to investigate more intelligent approaches. One strategy that suggests itself based on our study would involve learning the correct information to include in the headers based on historical data. Such a technique could then automatically suggest app modifications, in order to fix the ``buggy'' headers.

\subsection{Identifying Truly Redundant HTTP Requests}
\label{sec:sec:opportunity}

As discussed in Section~\ref{sec:rq}, redundant HTTP requests are good candidates for prefetching and caching. However, certain HTTP requests are only \emph{ostensibly redundant} in that they seem identical but actually yield different responses. Our final two research questions aim to shed light on this issue.
\begin{itemize}
\item \textbf{RQ$_8$} -- How prevalent are redundant HTTP requests?
\item \textbf{RQ$_9$} -- Are the identified ostensibly redundant requests truly redundant?

\end{itemize}
In our analysis, we have specifically focused on \texttt{GET} requests, as discussed previously.

\begin{figure}[b!]
	\centering
		\includegraphics[width=0.48\textwidth]{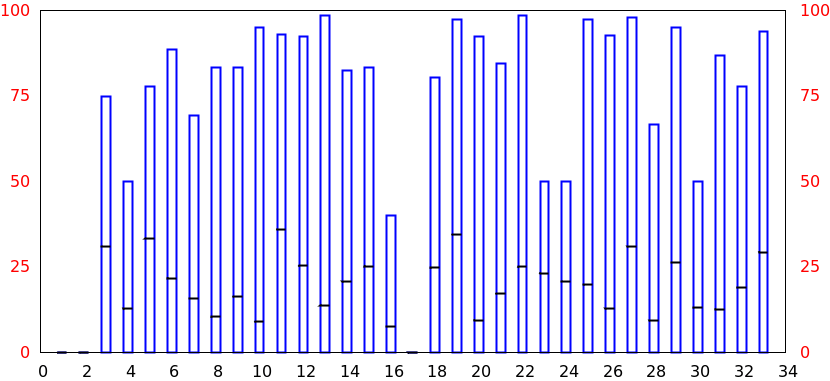}
	\caption{
    Minimum (bottom edges), maximum (top edges), and average (horizontal dashes) percentages of \emph{ostensibly} redundant requests in each app category.}
	\label{fig:redundant_percentage}
\end{figure}

\begin{figure}[b!]
	\centering
		\includegraphics[width=0.48\textwidth]{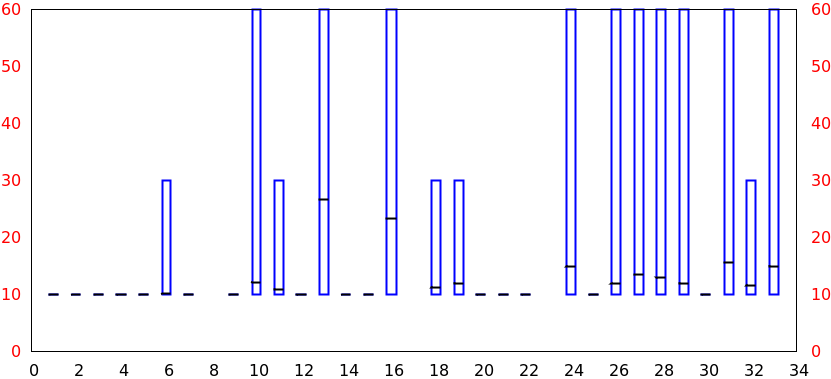}
	\caption{Minimum (bottom edges), maximum (top edges), and average (horizontal dashes) \emph{expiration times} for the redundant requests in each app category.}
	\label{fig:redundant_expire_time}
\end{figure}

\begin{figure}[b!]
	\centering
		\includegraphics[width=0.48\textwidth]{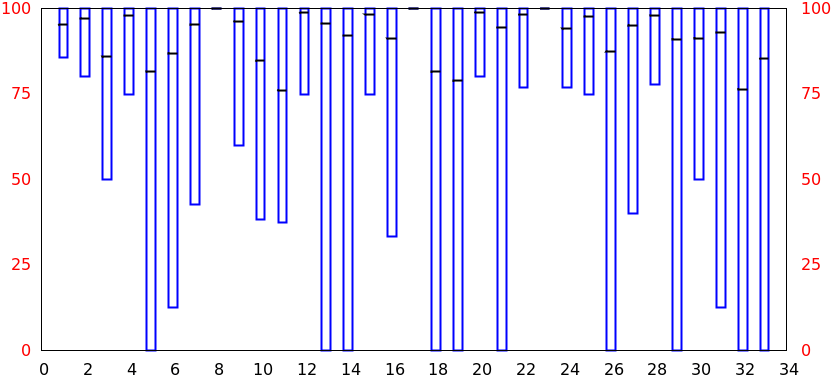}
	\caption{Minimum (bottom edges), maximum (top edges), and average (horizontal dashes) percentages of \emph{truly} redundant requests in each app category.}
	\label{fig:redundant_truly}
\end{figure}

To answer the above questions, upon completion of testing a given app (by executing the 3,000 events as explained in Section~\ref{sec:sec:testing}), we identify the ostensibly redundant requests in each app. 
We then run a script that executes the app by sending each identified request four times: at initial time $t$, $t+10$, $t+30$, and $t+60$ seconds. We check whether the responses change during this interval. This helps to identify HTTP requests that are truly redundant; the responses to those requests are thus suitable candidates for caching. 

Figure~\ref{fig:redundant_percentage} shows the minimum, maximum, and average percentages of the identified ostensibly redundant requests as compared to the total number of requests in each app category~(\emph{RQ$_8$}). Figure~\ref{fig:redundant_expire_time} shows the minimum, maximum, and average  expiration times for the identified requests~(\emph{RQ$_9$}). A request's expiration time is the time at which its response is different from the response received for the initial request at time $t$. Finally, Figure~\ref{fig:redundant_truly} shows the minimum, maximum, and average percentages of the \textit{truly} redundant requests~(\emph{RQ$_9$}).

As Figure~\ref{fig:redundant_percentage} shows, redundant requests comprise a significant proportion of all HTTP requests across most of the app categories. In certain apps, nearly 100\% of the requests are redundant, while the average across all apps is $\approx$20\%.
By themselves, these results would suggest considerable cacheability potential. 

This is further bolstered by some of the results in Figure~\ref{fig:redundant_expire_time}, which points to several apps in which the HTTP requests did not expire even after the full 60s. However, this is somewhat deceptive: The average request expiration time was 12s across the 33 app categories; it was exactly 10s for several of the categories; and only two categories---\emph{Food \& Drink} and \emph{House \& Home}---had average  expiration times over 20s. Since 10s was the shortest interval  used in our study, these results suggest that most redundant requests expire within a relatively short time period. This  should be taken into account when devising caching schemes for mobile apps. 

Finally, Figure~\ref{fig:redundant_truly} shows that, on average, an overwhelming majority of ostensibly redundant requests are \textit{truly redundant} across the 33 app categories. This means that the ostensibly redundant request did not expire at one or more of the 10s, 30s, and 60s checkpoints. In a number of individual apps, all ostensibly redundant requests are truly redundant (the maximum value of 100\%), while their average for app categories is as high as 92\%. This observation shows a large  opportunity for caching redundant requests in mobile apps.

\subsection{Implications}
\label{sec:sec:implications}
Our study provides evidence that prefetching and caching can be beneficial in a large number of mobile app scenarios. At the same time, we came across several apps in which prefetching and caching are unlikely to have 
significant, or any, benefits. At the least, these include the several hundred apps from our original set of subjects that only provide static content or make very few (1-3) HTTP requests. In fact, it is  possible that the number of these apps surpasses the 451 apps that do rely on the network and that we included in our final set of subjects  (recall the discussion in Section~\ref{sec:sec:final}). This outcome was at least somewhat surprising, given the long history of research on data prefetching and caching in distributed systems, of which mobile apps are only a more recent example.

A deeper analysis helps to identify several reasons behind this. 
For example, in hindsight it may have been expected that apps from the \emph{Libraries \& Demo} or \emph{Video Players \& Editors} categories provide static content, such as PDF viewers, organizers, digital books, and video players.
On the other hand, we did not expect to find almost as much static content in  \emph{Auto \& Vehicles}.
We already discussed in Section~\ref{sec:sec:final} that a number of apps from this category required login by suppling a license plate number. An additional, large number of apps also contained purely static content, such as instructions on how to perform car maintenance. This is reflected in our data: under 14\% of the \emph{Auto \& Vehicles} apps made it into our final set of 451 subjects (recall Table~\ref{tbl:category}). 

Another issue was presented by apps that used network communication that was either not based on HTTP or extensively used HTTP methods other than \texttt{GET}. For example, a number of apps in the \emph{Communications} category provide  instant messaging capabilities (including VoIP), while others actually implement browsers. \emph{Maps \& Navigation} provide GPS applications that differ significantly from typical HTTP services.
Yet another example are \emph{Finance} apps. Even though 44\% of these apps made it into our final set of subjects, a lot of them are banking apps that predominantly perform push-type operations, making them ill-suited for prefetching.

Even within the 451 final subject apps, there are clearly some  for which the benefits of prefetching and caching may be marginal. 
\emph{Games} presented an interesting case. Over $\frac{2}{3}$ of the apps in this category made it into our final set of subjects since they used sufficiently large numbers of HTTP requests. These apps also exhibited very high \texttt{Cache-Control} trustworthiness. On the other hand, as expected, their requests tended to expire very quickly and to have  little redundancy. Therefore, while a game app may be identified as a candidate for prefetching, the resulting cached data would become stale very quickly. In turn, this would possibly lead to incorrect app behavior or, just as bad, constant thrashing of the prefetching facilities that would cripple the app's performance.

These issues can be further illustrated with a somewhat crude analysis of an average app from our subject set. The average app sent 28 \texttt{GET} requests (recall Section~\ref{sec:sec:prefetchability}) as a result of the 3,000 automatically generated UI events. 20\% of those requests were truly redundant (recall Section~\ref{sec:sec:opportunity}). That means that up to 6 \texttt{GET} requests were prefetchable. Our previous work  PALOMA~\cite{paloma_icse} 
measured the processing of a single HTTP request to take slightly over 800ms under network conditions similar to ours. This would mean that an average app among our subjects would save only 4s by caching and reusing the results of the original request, assuming that the cache does not become stale.

While we must be cognizant of apps, such as those above, that are not especially amenable to prefetching and caching, several scenarios in our study paint a much more favorable picture. Consider the app from category 3 (\emph{Beauty}) that issued 1,243 \texttt{GET} requests (recall Figure~\ref{fig:get_number}), all of which are truly redundant (corresponding to the maximum value for app category 3 in Figure~\ref{fig:redundant_truly}). Even if we assume that the result of each redundant  request can only be reused once before it expires (recall from Figure~\ref{fig:redundant_expire_time} that the expiration time for app category 3 is 10s), that still yields 621 requests for which the results can be reused from the local cache. Assuming once again the same execution conditions as PALOMA's, this would result in massive execution-time savings, totaling 497s or 8.5 minutes. 

In summary, there is a notable opportunity for prefetching and caching in the mobile app domain. At the same time, the resulting techniques must take into account the characteristics of different app categories and different HTTP requests. Otherwise, the employed techniques may yield undesired outcomes, such as cache staleness, non-trivial performance overhead, and incorrect app behaviors.

\section{Threats to Validity}
\label{sec:threat}

Our study is based on top-ranked, free Android apps. Therefore, our results may not hold for paid apps or lower-ranked apps. However, over 90\% of the Android apps in the Google Play Store are free~\cite{freeapps}. Furthermore, top-ranked apps are used most widely. This suggests that our results should have broad applicability. 

We  excluded from our numerical analysis the apps that trigger fewer than four HTTP requests at runtime. However, part of the objective of our study was to explore this problem space. Specifically, we identified the reasons behind the apps' low numbers of requests (recall Section~\ref{sec:sec:final}). Furthermore, we acknowledged explicitly that the exclusion of these apps from the final set of subjects limits the applicability of our findings (recall Section~\ref{sec:sec:implications}). 


Our study is based on apps that use the HTTP protocol and two HTTP  libraries (\texttt{URLConnection} and \texttt{OkHttp}). Our findings are unlikely to be directly applicable to other protocols for network communication, and they may not carry over to other HTTP libraries. However, most mobile apps, and in particular Android apps, rely on HTTP~\cite{dai2013networkprofiler}. Furthermore, our focus is on the fundamental characteristics of HTTP requests and responses, and those characteristics do not change across different HTTP libraries. Including other libraries would naturally result in the inclusion of greater numbers of subject apps. However, given the popularity of the  HTTP libraries we selected, our results should be widely representative among Android apps.  

In our process for answering RQ$_{5}$, RQ$_{7}$, and RQ$_{9}$, we sent out sets of four  requests, at times $t$, $t+10$, $t+30$, and $t+60$ seconds (recall Sections~\ref{sec:sec:cachability} and~\ref{sec:sec:opportunity}).  
As shown in Figure~\ref{fig:redundant_expire_time}, redundant requests tend to expire at $t+10$ or soon thereafter. This indicates that $t+60$ is a sufficiently long period to identify truly redundant requests in most cases. Furthermore, mobile users tend to use an app for relatively short periods, so that prefetching and caching far in advance is not necessary and is likely to yield cache staleness. While choosing different time intervals would likely not lead to different results, finer-grained intervals may give us tighter bounds on request expiration times.

Finally, our app usage information was obtained via automated generation of UI events, as opposed to logging real user events. This may result in numbers and sequences of HTTP requests that are not representative of actual app use. However, the purpose of our study was to analyze all  possible HTTP requests that could be potentially triggered at runtime, and 3,000 random events were shown to be able to generate representative HTTP requests, as discussed in Section~\ref{sec:sec:testing}. Given the nature of the study and the large number of apps we aimed to analyze, it would have been unreasonable to attempt to find actual users for each app, while our results would potentially suffer from user-specific biases and idiosyncrasies in engaging the app. On the other hand, mimicking actual users with humans who are unfamiliar with the apps in question, which would have been a more likely alternative, would have suffered from the same potential problem as our automated testing. Furthermore, all of our research questions focus on individual HTTP requests rather than their sequences. Thus, real user traces would not lead to different results compared to random orders of runtime events. Finally, neither actual nor novice human users would have been able to repeatedly and reliably generate large numbers of events (3,000 per app execution in the main portion of our study, and up to 10,000 per execution in the preliminary analysis).

\vspace{-4mm}
\section{Related Work}
\label{sec:related_work}

Web prefetching and caching are entrenched techniques to reduce network latency since the Internet was born and have attracted a large body of work in browser domain, including measurement studies to understand web performance and identify performance bottlenecks~\cite{wang2011web,demystifying,qian2012web_ideal,erman2011cache,qian2013reduce,qian2014characterizing}, literature reviews and quantitative studies to compare fundamental prefetching and caching algorithms~\cite{webcachingsurvey,prefetchingimpact,wang2012far}, leveraging prefetching and caching techniques at different levels, such as studying user browsing behaviors~\cite{prefetchingclient,Lymberopoulos2012pocketweb}, providing API support for developers~\cite{mickens2010crom}, restructuring page load process~\cite{wang2016shandian,netravali2016polaris}, providing server or infrastructure support~\cite{rosen2017push,Ruamviboonsuk2017vroom,agababov2015flywheel,zhao2014cloud,butkiewicz2015klotski}.

The recent surge of mobile devices has attracted researchers to study prefetching and caching techniques in the context of mobile browsers and mobile apps. With the foundation of the traditional research in browser domain, mobile browser performance soon became a crowded research area~\cite{wang2012far,nejati2016depth,wang2011effective,wang2016shandian,Ruamviboonsuk2017vroom,demystifying}, but the research on mobile apps is still in its infancy. This is unfortunate because mobile users currently spend more than 80\% of their time in mobile apps rather than mobile browsers~\cite{appdominant}. In mobile app domain, Cachekeeper~\cite{zhang2013cachekeeper} studied the redundant HTTP traffic and proposed an OS-level caching service for HTTP requests on smartphones. PALOMA~\cite{paloma_icse} used program analysis to address ``what'' and ``when'' to prefetch certain HTTP requests in mobile apps. However, those techniques were only evaluated on a small number of apps and the performance depends on the flaws of web caching schemes employed in the original apps, thus it is not clear to what extent those techniques will be effective in practice. Those shortcomings of existing approaches motivated us to conduct an in-depth study that aims to understand the characteristics of HTTP requests in order to guide future research in mobile apps. Other existing works that focus on mobile app performance are complementary to our focus, such as pre-launching mobile apps~\cite{parate2013practical,yan2012fastlaunch,xu2013preference}, balancing Quality-of-Service (QoS) to suggest ``how much'' to prefetch~\cite{higgins2012informed,Baumann2017bytecount}, identifying performance bottlenecks~\cite{ravindranath2012appinsight}. 



\section{Conclusion}
\label{sec:conclusion}
In this paper, we presented the results of an extensive empirical study aimed at understanding the characteristics of HTTP requests and responses in mobile apps. We formulated nine research questions with the focus on the \textit{prefetchability} of  HTTP requests and \textit{cacheability} of  HTTP responses. Our overarching objective is to fill in the gap between the well-studied browser domain and comparatively less-explored mobile app domain, by motivating  and providing guidelines for future research in this area.  

Our results suggest that prefetching and caching can be useful across a wide range of mobile apps and scenarios, but they are not universally applicable and their benefits will vary. Certain app categories are more amenable for prefetching and caching. However, there is a non-trivial amount of variation even among different apps within a single cateogry. While our analysis reported in this paper does not provide definitive answers to questions of \emph{what}, \emph{when}, and \emph{how much} to prefetch/cache, it provides a process, tools, and data that form a foundation for answering those questions much more precisely than has been possible thus far.

\section*{Acknowledgment}
The authors thank William G.J. Halfond, Jiaping Gui, and the rest of their research group at the University of Southern California for providing us with the APKs for our subject apps. This work is supported by
the U.S. National Science Foundation under grants no. CCF-1618231
and CCF-1717963, U.S. Office of Naval Research under grant no.
N00014-17-1-2896, and by Huawei Technologies Co., Ltd.

\clearpage
\bibliographystyle{ACM-Reference-Format}
\balance
\bibliography{ase2018ref} 


\begin{thebibliography}{49}


\ifx \showCODEN    \undefined \def \showCODEN     #1{\unskip}     \fi
\ifx \showDOI      \undefined \def \showDOI       #1{#1}\fi
\ifx \showISBNx    \undefined \def \showISBNx     #1{\unskip}     \fi
\ifx \showISBNxiii \undefined \def \showISBNxiii  #1{\unskip}     \fi
\ifx \showISSN     \undefined \def \showISSN      #1{\unskip}     \fi
\ifx \showLCCN     \undefined \def \showLCCN      #1{\unskip}     \fi
\ifx \shownote     \undefined \def \shownote      #1{#1}          \fi
\ifx \showarticletitle \undefined \def \showarticletitle #1{#1}   \fi
\ifx \showURL      \undefined \def \showURL       {\relax}        \fi
\providecommand\bibfield[2]{#2}
\providecommand\bibinfo[2]{#2}
\providecommand\natexlab[1]{#1}
\providecommand\showeprint[2][]{arXiv:#2}

\bibitem[\protect\citeauthoryear{??}{adb}{2018}]%
        {adb}
 \bibinfo{year}{2018}\natexlab{}.
\newblock \bibinfo{title}{Android Debug Bridge}.
\newblock
\newblock
\urldef\tempurl%
\url{https://developer.android.com/studio/command-line/adb}
\showURL{%
\tempurl}


\bibitem[\protect\citeauthoryear{??}{fre}{2018}]%
        {freeapps}
 \bibinfo{year}{2018}\natexlab{}.
\newblock \bibinfo{title}{Distribution of free and paid Android apps in the
  Google Play Store}.
\newblock
\newblock
\urldef\tempurl%
\url{https://www.statista.com/statistics/266211/distribution-of-free-and-paid-android-apps/}
\showURL{%
\tempurl}


\bibitem[\protect\citeauthoryear{??}{nox}{2018}]%
        {noxplayer}
 \bibinfo{year}{2018}\natexlab{}.
\newblock \bibinfo{title}{NoxPlayer}.
\newblock
\newblock
\urldef\tempurl%
\url{https://www.bignox.com/}
\showURL{%
\tempurl}


\bibitem[\protect\citeauthoryear{??}{okh}{2018}]%
        {okhttp}
 \bibinfo{year}{2018}\natexlab{}.
\newblock \bibinfo{title}{OkHttp Documentation}.
\newblock
\newblock
\urldef\tempurl%
\url{http://square.github.io/okhttp/}
\showURL{%
\tempurl}


\bibitem[\protect\citeauthoryear{??}{ret}{2018}]%
        {retrofit}
 \bibinfo{year}{2018}\natexlab{}.
\newblock \bibinfo{title}{Retrofit Documentation}.
\newblock
\newblock
\urldef\tempurl%
\url{http://square.github.io/retrofit/}
\showURL{%
\tempurl}


\bibitem[\protect\citeauthoryear{??}{mon}{2018}]%
        {monkey}
 \bibinfo{year}{2018}\natexlab{}.
\newblock \bibinfo{title}{UI/Application Exerciser Monkey}.
\newblock
\newblock
\urldef\tempurl%
\url{https://developer.android.com/studio/test/monkey.html}
\showURL{%
\tempurl}


\bibitem[\protect\citeauthoryear{??}{url}{2018}]%
        {urlconnection}
 \bibinfo{year}{2018}\natexlab{}.
\newblock \bibinfo{title}{URLConnection Class Documentation}.
\newblock
\newblock
\urldef\tempurl%
\url{https://docs.oracle.com/javase/7/docs/api/java/net/URLConnection.html}
\showURL{%
\tempurl}


\bibitem[\protect\citeauthoryear{??}{vol}{2018}]%
        {volley}
 \bibinfo{year}{2018}\natexlab{}.
\newblock \bibinfo{title}{Volley overview}.
\newblock
\newblock
\urldef\tempurl%
\url{https://developer.android.com/training/volley/}
\showURL{%
\tempurl}


\bibitem[\protect\citeauthoryear{??}{emp}{2018}]%
        {empirical_website}
 \bibinfo{year}{2018}\natexlab{}.
\newblock \bibinfo{title}{The website of the raw data and the code of our
  analysis}.
\newblock
\newblock
\urldef\tempurl%
\url{https://github.com/felicitia/PALOMA-Analysis/tree/empirical}
\showURL{%
\tempurl}


\bibitem[\protect\citeauthoryear{Agababov, Buettner, Chudnovsky, Cogan,
  Greenstein, McDaniel, Piatek, Scott, Welsh, and Yin}{Agababov
  et~al\mbox{.}}{2015}]%
        {agababov2015flywheel}
\bibfield{author}{\bibinfo{person}{Victor Agababov}, \bibinfo{person}{Michael
  Buettner}, \bibinfo{person}{Victor Chudnovsky}, \bibinfo{person}{Mark Cogan},
  \bibinfo{person}{Ben Greenstein}, \bibinfo{person}{Shane McDaniel},
  \bibinfo{person}{Michael Piatek}, \bibinfo{person}{Colin Scott},
  \bibinfo{person}{Matt Welsh}, {and} \bibinfo{person}{Bolian Yin}.}
  \bibinfo{year}{2015}\natexlab{}.
\newblock \showarticletitle{Flywheel: Google's Data Compression Proxy for the
  Mobile Web.}. In \bibinfo{booktitle}{\emph{NSDI}}, Vol.~\bibinfo{volume}{15}.
  \bibinfo{pages}{367--380}.
\newblock


\bibitem[\protect\citeauthoryear{Baumann and Santini}{Baumann and
  Santini}{2017}]%
        {Baumann2017bytecount}
\bibfield{author}{\bibinfo{person}{Paul Baumann} {and} \bibinfo{person}{Silvia
  Santini}.} \bibinfo{year}{2017}\natexlab{}.
\newblock \showarticletitle{Every Byte Counts: Selective Prefetching for Mobile
  Applications}.
\newblock \bibinfo{journal}{\emph{Proc. ACM Interact. Mob. Wearable Ubiquitous
  Technol.}} \bibinfo{volume}{1}, \bibinfo{number}{2}, Article
  \bibinfo{articleno}{6} (\bibinfo{date}{June} \bibinfo{year}{2017}),
  \bibinfo{numpages}{29}~pages.
\newblock
\showISSN{2474-9567}
\urldef\tempurl%
\url{https://doi.org/10.1145/3090052}
\showDOI{\tempurl}


\bibitem[\protect\citeauthoryear{Bouras, Konidaris, and Kostoulas}{Bouras
  et~al\mbox{.}}{2004}]%
        {prefetchingimpact}
\bibfield{author}{\bibinfo{person}{Christos Bouras}, \bibinfo{person}{Agisilaos
  Konidaris}, {and} \bibinfo{person}{Dionysios Kostoulas}.}
  \bibinfo{year}{2004}\natexlab{}.
\newblock \showarticletitle{Predictive Prefetching on the Web and Its Potential
  Impact in the Wide Area}.
\newblock \bibinfo{journal}{\emph{World Wide Web}} \bibinfo{volume}{7},
  \bibinfo{number}{2} (\bibinfo{date}{June} \bibinfo{year}{2004}),
  \bibinfo{pages}{143--179}.
\newblock
\showISSN{1386-145X}
\urldef\tempurl%
\url{https://doi.org/10.1023/B:WWWJ.0000017208.87570.7a}
\showDOI{\tempurl}


\bibitem[\protect\citeauthoryear{Butkiewicz, Wang, Wu, Madhyastha, and
  Sekar}{Butkiewicz et~al\mbox{.}}{2015}]%
        {butkiewicz2015klotski}
\bibfield{author}{\bibinfo{person}{Michael Butkiewicz},
  \bibinfo{person}{Daimeng Wang}, \bibinfo{person}{Zhe Wu},
  \bibinfo{person}{Harsha~V Madhyastha}, {and} \bibinfo{person}{Vyas Sekar}.}
  \bibinfo{year}{2015}\natexlab{}.
\newblock \showarticletitle{Klotski: Reprioritizing Web Content to Improve User
  Experience on Mobile Devices.}. In \bibinfo{booktitle}{\emph{NSDI}},
  Vol.~\bibinfo{volume}{1}. \bibinfo{pages}{2--3}.
\newblock


\bibitem[\protect\citeauthoryear{Chaffey}{Chaffey}{2018}]%
        {appdominant}
\bibfield{author}{\bibinfo{person}{Dave Chaffey}.}
  \bibinfo{year}{2018}\natexlab{}.
\newblock \bibinfo{title}{Mobile Marketing Statistics compilation}.
\newblock
\newblock
\urldef\tempurl%
\url{https://www.smartinsights.com/mobile-marketing/mobile-marketing-analytics/mobile-marketing-statistics/}
\showURL{%
\tempurl}


\bibitem[\protect\citeauthoryear{Dai, Tongaonkar, Wang, Nucci, and Song}{Dai
  et~al\mbox{.}}{2013}]%
        {dai2013networkprofiler}
\bibfield{author}{\bibinfo{person}{Shuaifu Dai}, \bibinfo{person}{Alok
  Tongaonkar}, \bibinfo{person}{Xiaoyin Wang}, \bibinfo{person}{Antonio Nucci},
  {and} \bibinfo{person}{Dawn Song}.} \bibinfo{year}{2013}\natexlab{}.
\newblock \showarticletitle{Networkprofiler: Towards automatic fingerprinting
  of android apps}. In \bibinfo{booktitle}{\emph{INFOCOM, 2013 Proceedings
  IEEE}}. IEEE, \bibinfo{pages}{809--817}.
\newblock


\bibitem[\protect\citeauthoryear{Erman, Gerber, Hajiaghayi, Pei, Sen, and
  Spatscheck}{Erman et~al\mbox{.}}{2011}]%
        {erman2011cache}
\bibfield{author}{\bibinfo{person}{Jeffrey Erman}, \bibinfo{person}{Alexandre
  Gerber}, \bibinfo{person}{Mohammad Hajiaghayi}, \bibinfo{person}{Dan Pei},
  \bibinfo{person}{Subhabrata Sen}, {and} \bibinfo{person}{Oliver Spatscheck}.}
  \bibinfo{year}{2011}\natexlab{}.
\newblock \showarticletitle{To cache or not to cache: The 3G case}.
\newblock \bibinfo{journal}{\emph{IEEE Internet Computing}}
  \bibinfo{volume}{15}, \bibinfo{number}{2} (\bibinfo{year}{2011}),
  \bibinfo{pages}{27--34}.
\newblock


\bibitem[\protect\citeauthoryear{Fielding}{Fielding}{1999a}]%
        {rfc2616header}
\bibfield{author}{\bibinfo{person}{Roy Fielding}.}
  \bibinfo{year}{1999}\natexlab{a}.
\newblock \bibinfo{title}{RFC 2616, part of Hypertext Transfer Protocol --
  HTTP/1.1}.
\newblock
\newblock
\urldef\tempurl%
\url{https://www.w3.org/Protocols/rfc2616/rfc2616-sec14.html}
\showURL{%
\tempurl}


\bibitem[\protect\citeauthoryear{Fielding}{Fielding}{1999b}]%
        {rfc2616method}
\bibfield{author}{\bibinfo{person}{Roy Fielding}.}
  \bibinfo{year}{1999}\natexlab{b}.
\newblock \bibinfo{title}{RFC 2616, part of Hypertext Transfer Protocol --
  HTTP/1.1}.
\newblock
\newblock
\urldef\tempurl%
\url{https://tools.ietf.org/html/rfc2616#section-5.1.1}
\showURL{%
\tempurl}


\bibitem[\protect\citeauthoryear{Fielding}{Fielding}{1999c}]%
        {rfc2616get}
\bibfield{author}{\bibinfo{person}{Roy Fielding}.}
  \bibinfo{year}{1999}\natexlab{c}.
\newblock \bibinfo{title}{RFC 2616, part of Hypertext Transfer Protocol --
  HTTP/1.1}.
\newblock
\newblock
\urldef\tempurl%
\url{https://www.w3.org/Protocols/rfc2616/rfc2616-sec9.html}
\showURL{%
\tempurl}


\bibitem[\protect\citeauthoryear{Fielding}{Fielding}{1999d}]%
        {rfc2616safe}
\bibfield{author}{\bibinfo{person}{Roy Fielding}.}
  \bibinfo{year}{1999}\natexlab{d}.
\newblock \bibinfo{title}{RFC 2616, part of Hypertext Transfer Protocol --
  HTTP/1.1}.
\newblock
\newblock
\urldef\tempurl%
\url{https://tools.ietf.org/html/rfc2616#section-9}
\showURL{%
\tempurl}


\bibitem[\protect\citeauthoryear{Higgins, Flinn, Giuli, Noble, Peplin, and
  Watson}{Higgins et~al\mbox{.}}{2012}]%
        {higgins2012informed}
\bibfield{author}{\bibinfo{person}{Brett~D Higgins}, \bibinfo{person}{Jason
  Flinn}, \bibinfo{person}{Thomas~J Giuli}, \bibinfo{person}{Brian Noble},
  \bibinfo{person}{Christopher Peplin}, {and} \bibinfo{person}{David Watson}.}
  \bibinfo{year}{2012}\natexlab{}.
\newblock \showarticletitle{Informed mobile prefetching}. In
  \bibinfo{booktitle}{\emph{Proceedings of the 10th international conference on
  Mobile systems, applications, and services}}. ACM, \bibinfo{pages}{155--168}.
\newblock


\bibitem[\protect\citeauthoryear{KEMP}{KEMP}{2018}]%
        {wearesocial}
\bibfield{author}{\bibinfo{person}{SIMON KEMP}.}
  \bibinfo{year}{2018}\natexlab{}.
\newblock \bibinfo{title}{Digital in 2018}.
\newblock
\newblock
\urldef\tempurl%
\url{https://wearesocial.com/blog/2018/01/global-digital-report-2018}
\showURL{%
\tempurl}


\bibitem[\protect\citeauthoryear{Li, Lyu, Gui, and Halfond}{Li
  et~al\mbox{.}}{2016}]%
        {li2016automated}
\bibfield{author}{\bibinfo{person}{Ding Li}, \bibinfo{person}{Yingjun Lyu},
  \bibinfo{person}{Jiaping Gui}, {and} \bibinfo{person}{William~GJ Halfond}.}
  \bibinfo{year}{2016}\natexlab{}.
\newblock \showarticletitle{Automated energy optimization of http requests for
  mobile applications}. In \bibinfo{booktitle}{\emph{2016 IEEE/ACM 38th
  International Conference on Software Engineering (ICSE)}}. IEEE,
  \bibinfo{pages}{249--260}.
\newblock


\bibitem[\protect\citeauthoryear{Liu, Ma, Liu, Xie, and Huang}{Liu
  et~al\mbox{.}}{2016}]%
        {demystifying}
\bibfield{author}{\bibinfo{person}{X. Liu}, \bibinfo{person}{Y. Ma},
  \bibinfo{person}{Y. Liu}, \bibinfo{person}{T. Xie}, {and} \bibinfo{person}{G.
  Huang}.} \bibinfo{year}{2016}\natexlab{}.
\newblock \showarticletitle{Demystifying the Imperfect Client-Side Cache
  Performance of Mobile Web Browsing}.
\newblock \bibinfo{journal}{\emph{IEEE Transactions on Mobile Computing}}
  \bibinfo{volume}{15}, \bibinfo{number}{9} (\bibinfo{date}{Sept}
  \bibinfo{year}{2016}), \bibinfo{pages}{2206--2220}.
\newblock
\showISSN{1536-1233}
\urldef\tempurl%
\url{https://doi.org/10.1109/TMC.2015.2489202}
\showDOI{\tempurl}


\bibitem[\protect\citeauthoryear{Lymberopoulos, Riva, Strauss, Mittal, and
  Ntoulas}{Lymberopoulos et~al\mbox{.}}{2012}]%
        {Lymberopoulos2012pocketweb}
\bibfield{author}{\bibinfo{person}{Dimitrios Lymberopoulos},
  \bibinfo{person}{Oriana Riva}, \bibinfo{person}{Karin Strauss},
  \bibinfo{person}{Akshay Mittal}, {and} \bibinfo{person}{Alexandros Ntoulas}.}
  \bibinfo{year}{2012}\natexlab{}.
\newblock \showarticletitle{PocketWeb: Instant Web Browsing for Mobile
  Devices}. In \bibinfo{booktitle}{\emph{Proceedings of the Seventeenth
  International Conference on Architectural Support for Programming Languages
  and Operating Systems}} \emph{(\bibinfo{series}{ASPLOS XVII})}.
  \bibinfo{publisher}{ACM}, \bibinfo{address}{New York, NY, USA},
  \bibinfo{pages}{1--12}.
\newblock
\showISBNx{978-1-4503-0759-8}
\urldef\tempurl%
\url{https://doi.org/10.1145/2150976.2150978}
\showDOI{\tempurl}


\bibitem[\protect\citeauthoryear{Mickens, Elson, Howell, and Lorch}{Mickens
  et~al\mbox{.}}{2010}]%
        {mickens2010crom}
\bibfield{author}{\bibinfo{person}{James~W Mickens}, \bibinfo{person}{Jeremy
  Elson}, \bibinfo{person}{Jon Howell}, {and} \bibinfo{person}{Jay~R Lorch}.}
  \bibinfo{year}{2010}\natexlab{}.
\newblock \showarticletitle{Crom: Faster Web Browsing Using Speculative
  Execution.}. In \bibinfo{booktitle}{\emph{NSDI}}, Vol.~\bibinfo{volume}{10}.
  \bibinfo{pages}{9--9}.
\newblock


\bibitem[\protect\citeauthoryear{Nejati and Balasubramanian}{Nejati and
  Balasubramanian}{2016}]%
        {nejati2016depth}
\bibfield{author}{\bibinfo{person}{Javad Nejati} {and} \bibinfo{person}{Aruna
  Balasubramanian}.} \bibinfo{year}{2016}\natexlab{}.
\newblock \showarticletitle{An in-depth study of mobile browser performance}.
  In \bibinfo{booktitle}{\emph{Proceedings of the 25th International Conference
  on World Wide Web}}. International World Wide Web Conferences Steering
  Committee, \bibinfo{pages}{1305--1315}.
\newblock


\bibitem[\protect\citeauthoryear{Netravali, Goyal, Mickens, and
  Balakrishnan}{Netravali et~al\mbox{.}}{2016}]%
        {netravali2016polaris}
\bibfield{author}{\bibinfo{person}{Ravi Netravali}, \bibinfo{person}{Ameesh
  Goyal}, \bibinfo{person}{James Mickens}, {and} \bibinfo{person}{Hari
  Balakrishnan}.} \bibinfo{year}{2016}\natexlab{}.
\newblock \showarticletitle{Polaris: Faster Page Loads Using Fine-grained
  Dependency Tracking.}. In \bibinfo{booktitle}{\emph{NSDI}}.
  \bibinfo{pages}{123--136}.
\newblock


\bibitem[\protect\citeauthoryear{Parate, B{\"o}hmer, Chu, Ganesan, and
  Marlin}{Parate et~al\mbox{.}}{2013}]%
        {parate2013practical}
\bibfield{author}{\bibinfo{person}{Abhinav Parate}, \bibinfo{person}{Matthias
  B{\"o}hmer}, \bibinfo{person}{David Chu}, \bibinfo{person}{Deepak Ganesan},
  {and} \bibinfo{person}{Benjamin~M Marlin}.} \bibinfo{year}{2013}\natexlab{}.
\newblock \showarticletitle{Practical prediction and prefetch for faster access
  to applications on mobile phones}. In \bibinfo{booktitle}{\emph{Proceedings
  of the 2013 ACM international joint conference on Pervasive and ubiquitous
  computing}}. ACM, \bibinfo{pages}{275--284}.
\newblock


\bibitem[\protect\citeauthoryear{Qian, Huang, Erman, Mao, Sen, and
  Spatscheck}{Qian et~al\mbox{.}}{2013}]%
        {qian2013reduce}
\bibfield{author}{\bibinfo{person}{Feng Qian}, \bibinfo{person}{Junxian Huang},
  \bibinfo{person}{Jeffrey Erman}, \bibinfo{person}{Z~Morley Mao},
  \bibinfo{person}{Subhabrata Sen}, {and} \bibinfo{person}{Oliver Spatscheck}.}
  \bibinfo{year}{2013}\natexlab{}.
\newblock \showarticletitle{How to reduce smartphone traffic volume by 30\%?}.
  In \bibinfo{booktitle}{\emph{International Conference on Passive and Active
  Network Measurement}}. Springer, \bibinfo{pages}{42--52}.
\newblock


\bibitem[\protect\citeauthoryear{Qian, Quah, Huang, Erman, Gerber, Mao, Sen,
  and Spatscheck}{Qian et~al\mbox{.}}{2012}]%
        {qian2012web_ideal}
\bibfield{author}{\bibinfo{person}{Feng Qian}, \bibinfo{person}{Kee~Shen Quah},
  \bibinfo{person}{Junxian Huang}, \bibinfo{person}{Jeffrey Erman},
  \bibinfo{person}{Alexandre Gerber}, \bibinfo{person}{Zhuoqing Mao},
  \bibinfo{person}{Subhabrata Sen}, {and} \bibinfo{person}{Oliver Spatscheck}.}
  \bibinfo{year}{2012}\natexlab{}.
\newblock \showarticletitle{Web caching on smartphones: ideal vs. reality}. In
  \bibinfo{booktitle}{\emph{Proceedings of the 10th international conference on
  Mobile systems, applications, and services}}. ACM, \bibinfo{pages}{127--140}.
\newblock


\bibitem[\protect\citeauthoryear{Qian, Sen, and Spatscheck}{Qian
  et~al\mbox{.}}{2014}]%
        {qian2014characterizing}
\bibfield{author}{\bibinfo{person}{Feng Qian}, \bibinfo{person}{Subhabrata
  Sen}, {and} \bibinfo{person}{Oliver Spatscheck}.}
  \bibinfo{year}{2014}\natexlab{}.
\newblock \showarticletitle{Characterizing resource usage for mobile web
  browsing}. In \bibinfo{booktitle}{\emph{Proceedings of the 12th annual
  international conference on Mobile systems, applications, and services}}.
  ACM, \bibinfo{pages}{218--231}.
\newblock


\bibitem[\protect\citeauthoryear{Ravindranath, Padhye, Agarwal, Mahajan,
  Obermiller, and Shayandeh}{Ravindranath et~al\mbox{.}}{2012}]%
        {ravindranath2012appinsight}
\bibfield{author}{\bibinfo{person}{Lenin Ravindranath},
  \bibinfo{person}{Jitendra Padhye}, \bibinfo{person}{Sharad Agarwal},
  \bibinfo{person}{Ratul Mahajan}, \bibinfo{person}{Ian Obermiller}, {and}
  \bibinfo{person}{Shahin Shayandeh}.} \bibinfo{year}{2012}\natexlab{}.
\newblock \showarticletitle{AppInsight: Mobile App Performance Monitoring in
  the Wild.}. In \bibinfo{booktitle}{\emph{OSDI}}, Vol.~\bibinfo{volume}{12}.
  \bibinfo{pages}{107--120}.
\newblock


\bibitem[\protect\citeauthoryear{Rosen, Han, Hao, Mao, and Qian}{Rosen
  et~al\mbox{.}}{[n. d.]}]%
        {rosen2017push}
\bibfield{author}{\bibinfo{person}{Sanae Rosen}, \bibinfo{person}{Bo Han},
  \bibinfo{person}{Shuai Hao}, \bibinfo{person}{Z~Morley Mao}, {and}
  \bibinfo{person}{Feng Qian}.} \bibinfo{year}{[n. d.]}\natexlab{}.
\newblock \showarticletitle{Push or request: An investigation of http/2 server
  push for improving mobile performance}. In
  \bibinfo{booktitle}{\emph{Proceedings of the 26th International Conference on
  World Wide Web}}. \bibinfo{pages}{459--468}.
\newblock


\bibitem[\protect\citeauthoryear{Ruamviboonsuk, Netravali, Uluyol, and
  Madhyastha}{Ruamviboonsuk et~al\mbox{.}}{2017}]%
        {Ruamviboonsuk2017vroom}
\bibfield{author}{\bibinfo{person}{Vaspol Ruamviboonsuk}, \bibinfo{person}{Ravi
  Netravali}, \bibinfo{person}{Muhammed Uluyol}, {and}
  \bibinfo{person}{Harsha~V. Madhyastha}.} \bibinfo{year}{2017}\natexlab{}.
\newblock \showarticletitle{Vroom: Accelerating the Mobile Web with
  Server-Aided Dependency Resolution}. In \bibinfo{booktitle}{\emph{Proceedings
  of the Conference of the ACM Special Interest Group on Data Communication}}
  \emph{(\bibinfo{series}{SIGCOMM '17})}. \bibinfo{publisher}{ACM},
  \bibinfo{address}{New York, NY, USA}, \bibinfo{pages}{390--403}.
\newblock
\showISBNx{978-1-4503-4653-5}
\urldef\tempurl%
\url{https://doi.org/10.1145/3098822.3098851}
\showDOI{\tempurl}


\bibitem[\protect\citeauthoryear{Swaminathan and Raghavan}{Swaminathan and
  Raghavan}{2000}]%
        {prefetchingclient}
\bibfield{author}{\bibinfo{person}{N. Swaminathan} {and} \bibinfo{person}{S.~V.
  Raghavan}.} \bibinfo{year}{2000}\natexlab{}.
\newblock \showarticletitle{Intelligent prefetch in WWW using client behavior
  characterization}. In \bibinfo{booktitle}{\emph{Proceedings 8th International
  Symposium on Modeling, Analysis and Simulation of Computer and
  Telecommunication Systems (Cat. No.PR00728)}}. \bibinfo{pages}{13--19}.
\newblock
\showISSN{1526-7539}
\urldef\tempurl%
\url{https://doi.org/10.1109/MASCOT.2000.876424}
\showDOI{\tempurl}


\bibitem[\protect\citeauthoryear{Vall{\'e}e-Rai, Co, Gagnon, Hendren, Lam, and
  Sundaresan}{Vall{\'e}e-Rai et~al\mbox{.}}{1999}]%
        {soot}
\bibfield{author}{\bibinfo{person}{Raja Vall{\'e}e-Rai}, \bibinfo{person}{Phong
  Co}, \bibinfo{person}{Etienne Gagnon}, \bibinfo{person}{Laurie Hendren},
  \bibinfo{person}{Patrick Lam}, {and} \bibinfo{person}{Vijay Sundaresan}.}
  \bibinfo{year}{1999}\natexlab{}.
\newblock \showarticletitle{Soot - a Java Bytecode Optimization Framework}. In
  \bibinfo{booktitle}{\emph{Proceedings of the 1999 Conference of the Centre
  for Advanced Studies on Collaborative Research}}
  \emph{(\bibinfo{series}{CASCON '99})}. \bibinfo{publisher}{IBM Press},
  \bibinfo{pages}{13--}.
\newblock
\urldef\tempurl%
\url{http://dl.acm.org/citation.cfm?id=781995.782008}
\showURL{%
\tempurl}


\bibitem[\protect\citeauthoryear{Wang, Kong, Guo, and Chen}{Wang
  et~al\mbox{.}}{2013}]%
        {wang2013mobile}
\bibfield{author}{\bibinfo{person}{Haoyu Wang}, \bibinfo{person}{Junjun Kong},
  \bibinfo{person}{Yao Guo}, {and} \bibinfo{person}{Xiangqun Chen}.}
  \bibinfo{year}{2013}\natexlab{}.
\newblock \showarticletitle{Mobile web browser optimizations in the cloud era:
  A survey}. In \bibinfo{booktitle}{\emph{2013 IEEE 7th International Symposium
  on Service Oriented System Engineering (SOSE)}}. IEEE,
  \bibinfo{pages}{527--536}.
\newblock


\bibitem[\protect\citeauthoryear{Wang}{Wang}{1999}]%
        {webcachingsurvey}
\bibfield{author}{\bibinfo{person}{Jia Wang}.} \bibinfo{year}{1999}\natexlab{}.
\newblock \showarticletitle{A Survey of Web Caching Schemes for the Internet}.
\newblock \bibinfo{journal}{\emph{SIGCOMM Comput. Commun. Rev.}}
  \bibinfo{volume}{29}, \bibinfo{number}{5} (\bibinfo{date}{Oct.}
  \bibinfo{year}{1999}), \bibinfo{pages}{36--46}.
\newblock
\showISSN{0146-4833}
\urldef\tempurl%
\url{https://doi.org/10.1145/505696.505701}
\showDOI{\tempurl}


\bibitem[\protect\citeauthoryear{Wang, Krishnamurthy, and Wetherall}{Wang
  et~al\mbox{.}}{2016}]%
        {wang2016shandian}
\bibfield{author}{\bibinfo{person}{Xiao~Sophia Wang}, \bibinfo{person}{Arvind
  Krishnamurthy}, {and} \bibinfo{person}{David Wetherall}.}
  \bibinfo{year}{2016}\natexlab{}.
\newblock \showarticletitle{Speeding up Web Page Loads with Shandian.}. In
  \bibinfo{booktitle}{\emph{NSDI}}. \bibinfo{pages}{109--122}.
\newblock


\bibitem[\protect\citeauthoryear{Wang, Lin, Zhong, and Chishtie}{Wang
  et~al\mbox{.}}{2011a}]%
        {wang2011effective}
\bibfield{author}{\bibinfo{person}{Zhen Wang}, \bibinfo{person}{Felix~Xiaozhu
  Lin}, \bibinfo{person}{Lin Zhong}, {and} \bibinfo{person}{Mansoor Chishtie}.}
  \bibinfo{year}{2011}\natexlab{a}.
\newblock \showarticletitle{How effective is mobile browser cache?}. In
  \bibinfo{booktitle}{\emph{Proceedings of the 3rd ACM workshop on Wireless of
  the students, by the students, for the students}}. ACM,
  \bibinfo{pages}{17--20}.
\newblock


\bibitem[\protect\citeauthoryear{Wang, Lin, Zhong, and Chishtie}{Wang
  et~al\mbox{.}}{2011b}]%
        {wang2011web}
\bibfield{author}{\bibinfo{person}{Zhen Wang}, \bibinfo{person}{Felix~Xiaozhu
  Lin}, \bibinfo{person}{Lin Zhong}, {and} \bibinfo{person}{Mansoor Chishtie}.}
  \bibinfo{year}{2011}\natexlab{b}.
\newblock \showarticletitle{Why are web browsers slow on smartphones?}. In
  \bibinfo{booktitle}{\emph{Proceedings of the 12th Workshop on Mobile
  Computing Systems and Applications}}. ACM, \bibinfo{pages}{91--96}.
\newblock


\bibitem[\protect\citeauthoryear{Wang, Lin, Zhong, and Chishtie}{Wang
  et~al\mbox{.}}{2012}]%
        {wang2012far}
\bibfield{author}{\bibinfo{person}{Zhen Wang}, \bibinfo{person}{Felix~Xiaozhu
  Lin}, \bibinfo{person}{Lin Zhong}, {and} \bibinfo{person}{Mansoor Chishtie}.}
  \bibinfo{year}{2012}\natexlab{}.
\newblock \showarticletitle{How far can client-only solutions go for mobile
  browser speed?}. In \bibinfo{booktitle}{\emph{Proceedings of the 21st
  international conference on World Wide Web}}. ACM, \bibinfo{pages}{31--40}.
\newblock


\bibitem[\protect\citeauthoryear{Xu, Lin, Lu, Cardone, Lane, Chen, Campbell,
  and Choudhury}{Xu et~al\mbox{.}}{2013}]%
        {xu2013preference}
\bibfield{author}{\bibinfo{person}{Ye Xu}, \bibinfo{person}{Mu Lin},
  \bibinfo{person}{Hong Lu}, \bibinfo{person}{Giuseppe Cardone},
  \bibinfo{person}{Nicholas Lane}, \bibinfo{person}{Zhenyu Chen},
  \bibinfo{person}{Andrew Campbell}, {and} \bibinfo{person}{Tanzeem
  Choudhury}.} \bibinfo{year}{2013}\natexlab{}.
\newblock \showarticletitle{Preference, context and communities: a
  multi-faceted approach to predicting smartphone app usage patterns}. In
  \bibinfo{booktitle}{\emph{Proceedings of the 2013 International Symposium on
  Wearable Computers}}. ACM, \bibinfo{pages}{69--76}.
\newblock


\bibitem[\protect\citeauthoryear{Yan, Chu, Ganesan, Kansal, and Liu}{Yan
  et~al\mbox{.}}{2012}]%
        {yan2012fastlaunch}
\bibfield{author}{\bibinfo{person}{Tingxin Yan}, \bibinfo{person}{David Chu},
  \bibinfo{person}{Deepak Ganesan}, \bibinfo{person}{Aman Kansal}, {and}
  \bibinfo{person}{Jie Liu}.} \bibinfo{year}{2012}\natexlab{}.
\newblock \showarticletitle{Fast app launching for mobile devices using
  predictive user context}. In \bibinfo{booktitle}{\emph{Proceedings of the
  10th international conference on Mobile systems, applications, and
  services}}. ACM, \bibinfo{pages}{113--126}.
\newblock


\bibitem[\protect\citeauthoryear{Zhang, Tan, and Qun}{Zhang
  et~al\mbox{.}}{2013}]%
        {zhang2013cachekeeper}
\bibfield{author}{\bibinfo{person}{Yifan Zhang}, \bibinfo{person}{Chiu Tan},
  {and} \bibinfo{person}{Li Qun}.} \bibinfo{year}{2013}\natexlab{}.
\newblock \showarticletitle{CacheKeeper: a system-wide web caching service for
  smartphones}. In \bibinfo{booktitle}{\emph{Proceedings of the 2013 ACM
  international joint conference on Pervasive and ubiquitous computing}}. ACM,
  \bibinfo{pages}{265--274}.
\newblock


\bibitem[\protect\citeauthoryear{Zhao, Tak, and Cao}{Zhao
  et~al\mbox{.}}{2014}]%
        {zhao2014cloud}
\bibfield{author}{\bibinfo{person}{Bo Zhao}, \bibinfo{person}{Byung~Chul Tak},
  {and} \bibinfo{person}{Guohong Cao}.} \bibinfo{year}{2014}\natexlab{}.
\newblock \bibinfo{booktitle}{\emph{Mobile web browsing using the cloud}}.
\newblock \bibinfo{publisher}{Springer}.
\newblock


\bibitem[\protect\citeauthoryear{Zhao}{Zhao}{2017}]%
        {mobilesoft2017src}
\bibfield{author}{\bibinfo{person}{Y. Zhao}.} \bibinfo{year}{2017}\natexlab{}.
\newblock \showarticletitle{Toward Client-Centric Approaches for Latency
  Minimization in Mobile Applications}. In \bibinfo{booktitle}{\emph{2017
  IEEE/ACM 4th International Conference on Mobile Software Engineering and
  Systems (MOBILESoft)}}. \bibinfo{pages}{203--204}.
\newblock
\urldef\tempurl%
\url{https://doi.org/10.1109/MOBILESoft.2017.34}
\showDOI{\tempurl}


\bibitem[\protect\citeauthoryear{Zhao, Laser, Lyu, and Medvidovic}{Zhao
  et~al\mbox{.}}{2018}]%
        {paloma_icse}
\bibfield{author}{\bibinfo{person}{Yixue Zhao},
  \bibinfo{person}{Marcelo~Schmitt Laser}, \bibinfo{person}{Yingjun Lyu}, {and}
  \bibinfo{person}{Nenad Medvidovic}.} \bibinfo{year}{2018}\natexlab{}.
\newblock \showarticletitle{Leveraging Program Analysis to Reduce
  User-Perceived Latency in Mobile Applications}. In
  \bibinfo{booktitle}{\emph{Proceedings of the International Conference on
  Software Engineering (ICSE)}}.
\newblock


\end{thebibliography}
\end{document}